\documentclass[aps,pra,showpacs,twocolumn]{revtex4-1}
\usepackage{amssymb}
\usepackage{amsmath}
\usepackage{graphicx}
\usepackage{epsfig}

\begin{document}

\title{Topological Phase Transition Independent of System Non-Hermiticity}
\author{K. L. Zhang}
\author{H. C. Wu}
\author{L. Jin}
\email{jinliang@nankai.edu.cn}
\author{Z. Song}
\email{songtc@nankai.edu.cn}
\affiliation{School of Physics, Nankai University, Tianjin 300071, China}

\begin{abstract}
Non-Hermiticity can vary the topology of system, induce topological phase
transition, and even invalidate the conventional bulk-boundary
correspondence. Here, we show the introducing of non-Hermiticity without
affecting the topological properties of the original chiral symmetric
Hermitian systems. Conventional bulk-boundary correspondence holds,
topological phase transition and the (non)existence of edge states are
unchanged even though the energy bands are inseparable due to non-Hermitian
phase transition. Chern number for energy bands of the generalized
non-Hermitian system in two dimension is proved to be unchanged and
favorably coincides with the simulated topological charge pumping. Our
findings provide insights into the interplay between non-Hermiticity and
topology. Topological phase transition independent of non-Hermitian phase
transition is a unique feature that beneficial for future applications of
non-Hermitian topological materials.
\end{abstract}

\maketitle

\textit{Introduction.---}Parity-time ($\mathcal{PT}$) symmetry stimulates
the development of non-Hermitian physics \cite%
{BenderRPP,NM,FL,EL,Midya,YFChen,Christodoulides,MANiri}. Non-Hermitian
systems \cite{Ruschhaupt,ElOL,BPeng} exhibit many intriguing features and
applications that not limited to power oscillation \cite{Makris,Ruter},
coherent perfect absorption \cite{Chong}, unidirectionality behaviors \cite%
{Nat,FLNatMater,Ramezani14,PNAS,JLPRL}, single-mode laser \cite{LF,HH},
robust energy transfer \cite{Harris,SFan}, and exceptional point (EP)
enhanced sensing \cite{Wiersig,ZPLiu,WChen,HHodaei} due to its nonorthogonal
eigenstates and the exotic topology of EPs \cite%
{Heiss01,BerryMailybaev,TGao,CTChanPRX,Doppler,LJin18}. The scope of
topological phase of matter has also been extended to non-Hermitian region
\cite%
{Rudner,Szameit11,Hughes,Esaki,Diehl,GQLiang,Malzard,Kartashov,YFChenPNAS,SLieu,MPan,HS,RFleury,CerjanPRB,Kunst1812,XWLuo18,Cancellieri,YXu19}
and stimulates several interesting discussions on $\mathcal{PT}$-symmetric
topological interface states \cite%
{OL,Poli,Weimann,Menke,LJin17,PXue,Yuce,Ghatak,KawabataPRB98,XNi,LJLPRB,SLonghi19}%
, non-Hermitian bands theory \cite{Shen,Murakami}, topological invariants
\cite{Rakovszky,Shen,YXu,Leykam,SLin,WRPRA,ZXZ19,HJiang}, EP lines and
surfaces \cite{BZhen,ZhouScience,YXu,CerjanEPring}, semimetals \cite%
{Molina,Carllstrom,CHLeeTidal,Zyuzin,Schmidt,JHu,HZhang}, high-order
topological phases \cite{TLiu,Ezawa,JGong,Edvardsson,XWLuo}, and symmetry
protected non-Hermitian topological phases \cite{SLin,Budich,Yoshida,Okugawa}%
. Topological classification are discussed for general non-Hermitian systems~%
\cite{ZGong,HZhou,Kawabata1812,KawabataNC}, for non-Hermitian systems with
reflection symmetry \cite{CHLiu}, and alteratively classified by the
geometric features of singularity ring \cite{LinhuLi}. The non-Hermiticity
and non-Abelian gauge potentials can create interesting topological phases
\cite{JCai}. The robust and efficient topological edge state lasing is an
interesting application of non-Hermitian topological systems~\cite%
{BBahari,PStJean,HZhao,MParto,Science,Kartashov19}.

Topological invariant constructed from the bulk system predicts the
topological phase transition and the (non)existence of edge states in the
system under open boundary condition \cite{Kane}, this is referred to as the
conventional bulk-boundary correspondence (CBBC). In certain non-Hermitian
topological systems, the bulk topology fails to predict the edge states and
topological phase transition in systems under open boundary condition \cite%
{TonyPRL,Martinez,KawabataPRB,CHLee,LJL}; nevertheless, the exotic
bulk-boundary correspondences have been reported \cite%
{ZGong,ZWang,WYi,Kunst,Herviou}. A non-Bloch bulk Hamiltonian is constructed
to resolve this issue \cite{ZWang}; alternatively, topological invariant is
established from the biorthogonal edge modes \cite{Kunst,Edvardsson}.
Furthermore, chiral inversion symmetry is uncovered to protect the CBBC in
non-Hermitian topological systems~\cite{JLBBC}, and the CBBC and the skin
modes are elucidated in the viewpoint of non-Hermitian Aharonov-Bohm effect;
alternatively, they are elucidated from a transfer matrix perspective \cite%
{FKK} and the Green's function method \cite{Dan}. Besides, the
non-Hermiticity can solely induce topological phase, which has been
demonstrated in trivial Hermitian systems associated with staggered gain and
loss \cite{Takata}, asymmetric coupling amplitude \cite{JLBBC}, and
imaginary coupling \cite{KawabataNC}, respectively. Therefore, the
non-Hermiticity can alter the topology of system, induce topological phase
transition, and even ruin the CBBC; in contrast to the topology changed by
non-Hermiticity, retaining the topology of Hermitian system in the
non-Hermitian generalization is a critical and meaningful challenge for
non-Hermitian topological phase of matter.

In this work, we systematically elucidate the introducing of non-Hermiticity
\textit{without} altering the topological phase transition in the original
chiral symmetric Hermitian system; the proposed non-Hermitian topological
system holds the CBBC and shares identical topological properties including
the (non)existence of topologically protected edge states with their parent
Hermitian system, even though the energy bands are deformed into the complex
domain and inseparable. The complete set of eigenstates of the non-Hermitian
system is exactly mapped from the eigenstates of the original Hermitian
system by a set of local transformations; the mapping allows direct
projections of their geometric quantities. In the non-Hermitian
generalization, five symmetry classes with chiral symmetry are mapped to the
other five symmetry classes without chiral symmetry, respectively; the Chern
number in two dimension (2D) is proved to be unchanged, and the numerically
simulated topological charge pumping favorably agrees with the Chern number.

\textit{Mapping the topology.---}Chiral symmetric systems can be written in
the block off-diagonal form \cite{Ryu}%
\begin{equation}
H=\left(
\begin{array}{cc}
0 & D \\
D^{\dag } & 0%
\end{array}%
\right) ,  \label{hq}
\end{equation}%
where we consider $D$\ as an arbitrary $n\times n$\ matrix. The basis in $H$
can be any degree of freedom, such as the real space coordinate, spin, or
other orthogonal complete set. $H$ is referred to as the original Hermitian
Hamiltonian, from which a non-Hermitian Hamiltonian $\mathcal{H}$ is created%
\begin{equation}
\mathcal{H}=H+i\gamma \sigma _{z}\otimes I_{n},  \label{NH h}
\end{equation}%
where $\sigma _{z}$ is the Pauli matrix, and $I_{n}$ denotes the $n\times n$
identity matrix. The non-Hermitian term $i\gamma $ in $\mathcal{H}$ does not
only play the role of on-site potential, but also the non-Hermitian hopping
with asymmetric amplitudes in the real space. For chiral symmetric systems
not in the bipartite lattice form, taken the Creutz ladder as an
illustration \cite{TonyPRL}, the gain and loss introduced in the block
off-diagonal form of $H$ [Eq. (\ref{hq})] are equivalent to introducing
asymmetric couplings in the ladder legs~\cite{JLBBC}. In addition,
introducing non-Hermiticity breaks the chiral symmetry.

Equation (\ref{NH h}) provides a way of non-Hermitian generalization without
altering the topological phase transition in original Hermitian systems. To
characterize the topological properties, we consider the Hamiltonian $H(%
\mathbf{k)}$ in the momentum space, which is the core matrix of a Bloch or a
BdG system \cite{Ryu}. $\mathbf{k}$ is the momentum and all the information
of the topological system is encoded in $H(\mathbf{k)}$. The Schr\"{o}dinger
equation is $H(\mathbf{k})\left\vert \phi _{\lambda }^{\rho }(\mathbf{k}%
)\right\rangle =\varepsilon _{\lambda }^{\rho }(\mathbf{k})\left\vert \phi
_{\lambda }^{\rho }(\mathbf{k})\right\rangle $ and $\mathcal{H}(\mathbf{k}%
)\left\vert \varphi _{\lambda }^{\rho }(\mathbf{k})\right\rangle =\epsilon
_{\lambda }^{\rho }(\mathbf{k})\left\vert \varphi _{\lambda }^{\rho }(%
\mathbf{k})\right\rangle $, where $\rho =\pm $ represents the upper or lower
energy band, and $\lambda \in \lbrack 1,l]$\ denotes the band index,
assuming the total number of energy bands $2l$. The eigenenergy of $\mathcal{%
H}(\mathbf{k})$ is%
\begin{equation}
\epsilon _{\lambda }^{\rho }\left( \mathbf{k}\right) =\rho \lbrack
\varepsilon _{\lambda }^{\rho }\left( \mathbf{k}\right) ^{2}-\gamma
^{2}]^{1/2}.  \label{Energy}
\end{equation}%
The energy is either real or imaginary. The eigenstate of $\mathcal{H}(%
\mathbf{k})$ for eigenenergy $\epsilon _{\lambda }^{\rho }(\mathbf{k})$\ is
obtained through a mapping (see Appendix A)%
\begin{equation}
\left\vert \varphi _{\lambda }^{\rho }(\mathbf{k})\right\rangle =M_{\lambda
}^{\rho }(\mathbf{k},\gamma )\left\vert \phi _{\lambda }^{\rho }(\mathbf{k}%
)\right\rangle .  \label{State}
\end{equation}%
with the mapping matrix
\begin{equation}
M_{\lambda }^{\rho }(\mathbf{k},\gamma )=\left(
\begin{array}{cc}
a_{\lambda }^{\rho }\left( \mathbf{k},\gamma \right) I_{n} & 0 \\
0 & I_{n}%
\end{array}%
\right) ,  \label{Mapping}
\end{equation}%
where $a_{\lambda }^{\rho }\left( \mathbf{k},\gamma \right) =[\epsilon
_{\lambda }^{\rho }(\mathbf{k})+i\gamma ]/\varepsilon _{\lambda }^{\rho }(%
\mathbf{k})$ is a unit modulus complex number for real $\epsilon _{\lambda
}^{\rho }\left( \mathbf{k}\right) $, and $\left\vert a_{\lambda }^{\rho
}\left( \mathbf{k},\gamma \right) \right\vert \neq 1$ for imaginary $%
\epsilon _{\lambda }^{\rho }\left( \mathbf{k}\right) $. The mapping $%
M_{\lambda }^{\rho }(\mathbf{k,}\gamma )$ acts as a local transformation,
which is essential for the inheritance of topological features from the
original Hermitian system in the non-Hermitian generalization.

\textit{Bulk-boundary correspondence.---}CBBC does not always hold in
non-Hermitian topological systems \cite{TonyPRL,Martinez,ZGong,ZWang,WYi,LJL}%
, where effective imaginary gauge field induces a non-Hermitian
Aharonov-Bohm effect that invalidates the CBBC \cite{CHLee,JHu,ZWang,JLBBC}.
Considering a bipartite lattice Hamiltonian \cite{Lieb}, the gain and loss
are respectively introduced in two sublattices for the proposed manner [Eq.~(%
\ref{NH h})]. $\mathcal{H}$ is a $2n$-site non-Hermitian lattice constituted
by $n$ coupled $\mathcal{PT}$-symmetric dimers. Applying a unitary
transformation, the intra dimer coupling $J$ associated with the gain and
loss $\gamma $ in a $\mathcal{PT}$-symmetric dimer changes into the
asymmetric intra dimer couplings $J\pm \gamma $ (see Appendix B), which
appears as inter sublattice couplings; and the effective imaginary gauge
field is absent along the translational invariant direction of the
sublattice, the non-Hermitian Aharonov-Bohm effect does not occur, and the
CBBC holds.

Alternatively, the validity of CBBC can be straightforwardly understood from
the mapping between the original Hermitian topological system and the
generalized non-Hermitian topological system. Although the energy bands are
tightened [Eq.~(\ref{Energy})] after introduced the non-Hermiticity, the
band structures and their topologies are unchanged. The mapping matrix [Eq.~(%
\ref{Mapping})] retains the profile of the eigenstates; the Dirac
probability distribution of the eigenstates inside each sublattice is
unchanged after mapping [Eq.~(\ref{State})]. The CBBC is valid for the
non-Hermitian system $\mathcal{H}$ and \textit{does not} require the
symmetry protection, this differs from that in Ref. \cite{JLBBC}.

\textit{Mapping of symmetry classes.---}In the ten Altland-Zirnbauer classes
\cite{AZ}, topological systems with chiral symmetry include five symmetry
classes and satisfy $\mathcal{S}H(\mathbf{k)}\mathcal{S}^{-1}=-H(\mathbf{k)}$
\cite{Ryu}, where the chiral operator is $\mathcal{S}=\sigma _{z}\otimes
I_{n}\mathcal{=S}^{-1}$. The symmetry class $\mathrm{AIII}$ does not have
additional discrete symmetries, only a combined time-reversal ($\mathcal{T}$%
) and particle-hole ($\mathcal{C}$) symmetry $\mathcal{S}=\mathcal{TC}$ is
present. The symmetry classes $\mathrm{BDI}$, $\mathrm{CII}$, $\mathrm{CI}$,
and $\mathrm{DIII}$ have additional time-reversal and particle-hole
symmetries under $\mathcal{T}H(-\mathbf{k})\mathcal{T}^{-1}=H(\mathbf{k)}$
and $\mathcal{C}H(-\mathbf{k)}\mathcal{C}^{-1}=-H(\mathbf{k)}$. After
introducing the non-Hermiticity, the chiral symmetry vanishes in
non-Hermitian topological systems $\mathcal{H(}\mathbf{k)}=H\mathcal{(}%
\mathbf{k)}+i\gamma \sigma _{z}\otimes I_{n}$.

From $\mathcal{S}=\mathcal{TC}$, we obtain $\mathcal{T}^{-1}=\mathcal{CS}%
^{-1}=\mathcal{CS}$. Then we have $\mathcal{T}\left( i\gamma \mathcal{S}%
\right) \mathcal{T}^{-1}=-i\gamma \mathcal{T}\left( \mathcal{TC}\right)
\left( \mathcal{CS}\right) =-i\gamma \mathcal{T}^{2}\mathcal{C}^{2}\mathcal{%
\mathcal{S}}$. From $\mathcal{T}=\mathcal{SC}^{-1}$, we obtain $\mathcal{C}%
\left( i\gamma \mathcal{S}\right) \mathcal{C}^{-1}=-i\gamma \mathcal{CT}^{2}%
\mathcal{\mathcal{T}}^{-1}=-i\gamma \mathcal{T}^{2}\mathcal{C}\left(
\mathcal{C\mathcal{S}}\right) =-i\gamma \mathcal{T}^{2}\mathcal{C}^{2}%
\mathcal{\mathcal{S}}$. Thus under the action of time-reversal and
particle-hole operators, the non-Hermitian term $i\gamma (\sigma _{z}\otimes
I_{n})=i\gamma \mathcal{S}$ satisfies
\begin{equation}
\mathcal{T}(i\gamma \mathcal{S})\mathcal{T}^{-1}=\mathcal{C}(i\gamma
\mathcal{S})\mathcal{C}^{-1}=-i\gamma \mathcal{T}^{2}\mathcal{\mathcal{C}}%
^{2}\mathcal{S}.
\end{equation}

After the non-Hermitian generalization, the symmetry class $\mathrm{AIII}$ $H%
\mathcal{(}\mathbf{k)}$ changes to symmetry class $\mathrm{A}$ $\mathcal{H(}%
\mathbf{k)}$. The chiral orthogonal ($\mathrm{BDI)}$ class has $\mathcal{T}%
^{2}\mathcal{\mathcal{C}}^{2}\mathcal{S}=\left( +1\right) \left( +1\right)
\mathcal{S}$. Thus,
\begin{equation}
\mathcal{TH(}\mathbf{k)}\mathcal{T}^{-1}=H(-\mathbf{k)}-i\gamma (\sigma
_{z}\otimes I_{n})\neq \mathcal{H}(-\mathbf{k),}
\end{equation}
but
\begin{equation}
\mathcal{CH}\mathbf{(k)}\mathcal{C}^{-1}=-H(-\mathbf{k)}-i\gamma (\sigma
_{z}\otimes I_{n})=-\mathcal{H}(-\mathbf{k}).
\end{equation}
The time-reversal symmetry breaks, but the particle-hole symmetry holds for
the non-Hermitian topological systems; the symmetry class $\mathrm{BDI}$ is
mapped to the symmetry class $\mathrm{D}$. The chiral symplectic ($\mathrm{%
CII}$)\ class has $\mathcal{T}^{2}\mathcal{\mathcal{C}}^{2}\mathcal{S}%
=\left( -1\right) \left( -1\right) \mathcal{S}$; similarly, only the
particle-hole symmetry holds\textbf{\ }$\mathcal{CH}(-\mathbf{k})\mathcal{C}%
^{-1}=-\mathcal{H}\mathbf{(k})$ and the symmetry class $\mathrm{CII}$ is
mapped to the symmetry class $\mathrm{C}$. For the other two symmetry
classes with chiral symmetry, the symmetry class $\mathrm{CI}$ has $\mathcal{%
T}^{2}\mathcal{\mathcal{C}}^{2}\mathcal{S}=\left( +1\right) \left( -1\right)
\mathcal{S}$ and the symmetry class $\mathrm{DIII}$ has $\mathcal{T}^{2}%
\mathcal{\mathcal{C}}^{2}\mathcal{S}=\left( -1\right) \left( +1\right)
\mathcal{S}$; both two classes satisfy
\begin{equation}
\mathcal{TH}\mathbf{(k)}\mathcal{T}^{-1}=H(-\mathbf{k)}+i\gamma (\sigma
_{z}\otimes I_{n})=\mathcal{H}(-\mathbf{k),}
\end{equation}
but
\begin{equation}
\mathcal{CH}(\mathbf{k)}\mathcal{C}^{-1}=-H(-\mathbf{k)}+i\gamma (\sigma
_{z}\otimes I_{n})\neq -\mathcal{H}(-\mathbf{k).}
\end{equation}
The time-reversal symmetry holds, but the particle-hole symmetry breaks in
the non-Hermitian generalization. The mappings of symmetry classes are $%
\mathrm{CI}\rightarrow \mathrm{AI}$ and $\mathrm{DIII}\rightarrow \mathrm{AII%
}$. In summary, the introduced non-Hermiticity breaks the chiral symmetry
and one of the time-reversal and particle-hole symmetries; the five symmetry
classes with chiral symmetry shift to the other five symmetry classes
without chiral symmetry
\begin{equation}
\mathrm{AIII\rightarrow A,BDI\rightarrow D,CII\rightarrow C,CI\rightarrow
AI,DIII\rightarrow AII.}
\end{equation}

\textit{Chern number in 2D systems.---}Considering a 2D topological system,
the (first) Chern numbers for each band of the two Hamiltonians $H(\mathbf{k}%
)$ and $\mathcal{H}(\mathbf{k})$ are exactly identical. In the absence of
EPs, the energy bands are separable, the four types of Chern numbers defined
under the right and left eigenstates of $\mathcal{H}(\mathbf{k})$ are
identical \cite{Shen,YXu19} (see Appendix C). For separable bands of the
Hermitian Hamiltonian, the bands of non-Hermitian Hamiltonian are
\textquotedblleft separable" in practice even if the energy bands merge in
the presence of EPs. To see that the Chern number is a topological invariant
and does not change in the mapping, we employ the conventional definition;
unlike the lack of biorthonormal basis at EPs \cite{Ali02}, whose
biorthonormal probability vanishes at the exceptional point $\mathbf{k}$ for
certain bands, the Berry connections $\mathbf{A}_{\lambda }^{\rho
}=i\left\langle \phi _{\lambda }^{\rho }(\mathbf{k})\right\vert \nabla _{%
\mathbf{k}}\left\vert \phi _{\lambda }^{\rho }(\mathbf{k})\right\rangle $
for $H(\mathbf{k})$ and $\overrightarrow{\mathcal{A}}_{\lambda }^{\rho
}=i\left\langle \varphi _{\lambda }^{\rho }(\mathbf{k})\right\vert \nabla _{%
\mathbf{k}}\left\vert \varphi _{\lambda }^{\rho }(\mathbf{k})\right\rangle $
for $\mathcal{H}(\mathbf{k})$ based on the right eigenstates are always
well-defined. The nabla operator is $\nabla _{\mathbf{k}}=(\partial
_{k_{x}},\partial _{k_{y}})$.

Direct derivation yields $\overrightarrow{\mathcal{A}}_{\lambda }^{\rho }=%
\mathbf{A}_{\lambda }^{\rho }-\left( 1/2\right) \nabla _{\mathbf{k}%
}\vartheta $ [$\overrightarrow{\mathcal{A}}_{\lambda }^{\rho }=\mathbf{A}%
_{\lambda }^{\rho }+$ $\left\langle \phi _{\lambda }^{-\rho }(\mathbf{k}%
)\right\vert i\nabla _{\mathbf{k}}\left\vert \phi _{\lambda }^{\rho }(%
\mathbf{k})\right\rangle \epsilon _{\lambda }^{\rho }(\mathbf{k})/\left(
i\gamma \right) $] for real (imaginary) spectrum, in which $\vartheta
=\arctan [\gamma /\epsilon _{\lambda }^{\rho }(\mathbf{k})]$; the relation
is gauge dependent (see Appendix C), however, the Berry curvatures are gauge
independent\ $\mathbf{B}_{\lambda }^{\rho }=\nabla _{\mathbf{k}}\times
\mathbf{A}_{\lambda }^{\rho }$, $\overrightarrow{\mathcal{B}}_{\lambda
}^{\rho }=\nabla _{\mathbf{k}}\times \overrightarrow{\mathcal{A}}_{\lambda
}^{\rho }$, and obey $\mathbf{B}_{\lambda }^{\rho }=\overrightarrow{\mathcal{%
B}}_{\lambda }^{\rho }$ ($\mathbf{B}_{\lambda }^{\rho }\neq \overrightarrow{%
\mathcal{B}}_{\lambda }^{\rho }$) for real (imaginary) spectrum; and the
contribution of the later term in $\overrightarrow{\mathcal{A}}_{\lambda
}^{\rho }$ for the Chern number $c_{\lambda }^{\rho }$ is zero,%
\begin{equation}
c_{\lambda }^{\rho }=\frac{1}{2\pi }\oint \mathbf{B}_{\lambda }^{\rho }d^{2}%
\mathbf{k=}\frac{1}{2\pi }\oint \overrightarrow{\mathcal{B}}_{\lambda
}^{\rho }d^{2}\mathbf{k.}  \label{Chern n}
\end{equation}%
This is referred to as the topological invariant mapping. Notably, the
mapping Eq.~(\ref{State}) is directly applicable to the \textit{edge states}%
. These conclusions are \textit{not} relevant to the reality of energy bands
and the presence of EPs. The Chern numbers for each energy band of both
systems are identical even if the bands merge in the presence of EPs (see
Appendix C).

Ultracold atomic gases \cite{Goldman,Cooper}, acoustic lattices \cite%
{RF,YFChenNP}, electrical circuits \cite{EzawaPRB,RYu,CHLeeCP}, and various
microwave, optical, and photonic systems \cite{Hafezi,LLu,Ozawa} have became
fertile platforms for studying topological phase of matter. Through
introducing additional losses, passive non-Hermitian topological systems are
created \cite{CerjanEPring,ZhouScience}; the properties of $\mathcal{PT}$%
-symmetric systems with balanced gain and loss are exacted from the passive
systems by shifting a common loss rate. Nowadays, the non-Hermitian
topological systems are experimentally realized via sticking absorbers in
the dielectric resonator array \cite{Poli}, cutting the waveguides in
coupled optical waveguide lattice \cite{CerjanEPring}, and fabricating the
radiative loss in open systems of photonic crystals \cite%
{Weimann,ZhouScience}. Active elements are required to realize robust
topological edge state lasing \cite%
{BBahari,Science,HZhao,PStJean,MParto,Kartashov19}, where external pumping
is implemented to acquire the gain. The prototypical non-Hermitian
topological system is the 1D complex Su--Schrieffer--Heeger (SSH) model \cite%
{OL}; here we consider a simple extension to interpret the Chern number in
the non-Hermitian generalization from the viewpoint of topological charge
pumping. Figure~\ref{fig1}(a) shows the 1D comb lattice formed via staggered
side-coupled additional sites to the intensively investigated complex SSH
lattice in experiment \cite{Poli,Weimann,MPan,MParto,PStJean,HZhao,YDChong18}%
.

\textit{Topological charge pumping}.---In the momentum space, the core
matrix is%
\begin{equation}
H\left( k\right) =\left(
\begin{array}{cccc}
0 & 0 & \mu _{k} & \kappa _{+} \\
0 & 0 & \kappa _{-} & 0 \\
\mu _{-k} & \kappa _{-} & 0 & 0 \\
\kappa _{+} & 0 & 0 & 0%
\end{array}%
\right) ,
\end{equation}%
where $\mu _{k}=J(1-\delta )+J(1+\delta )e^{ik}$ and the system parameters
are $\delta =\delta _{0}+R\cos \theta $ and $\kappa _{\pm }=\kappa _{0}\pm
(1/2)R\sin \theta $ (set $\kappa \equiv \kappa _{+}-\kappa _{-}=R\sin \theta
$), forming a loop $L$ with radius $R$($>0$)\ in the parameter space. The
core matrix of the non-Hermitian generalization $\mathcal{H}\left( k\right) $
gives%
\begin{equation}
\mathcal{H}(k)=H(k)+i\gamma \sigma _{z}\otimes I_{2}.
\end{equation}%
$H(k)$ belongs to symmetry class \textrm{BDI}, and $\mathcal{H}(k)$ belongs
to symmetry class \textrm{D} only with the particle-hole symmetry, $\mathcal{%
CH}(k)\mathcal{C}^{-1}=-\mathcal{H}(-k)$, where $\mathcal{C}=\left( \sigma
_{z}\otimes I_{2}\right) \mathcal{K}$, $I_{2}$ is the $2\times 2$ identity
matrix, and $\mathcal{K}$ is the complex conjugation. Eigenstates $%
\{|\varphi _{\lambda }^{\rho }(k)\rangle ,|\eta _{\lambda }^{\rho
}(k)\rangle \}$\ for $\{\mathcal{H}(k),\mathcal{H}^{\dagger }(k)\}$ are
obtained from the eigenstates of $H(k)$ through mapping (see Appendix D).
The corresponding energy for eigenstate $\left\vert \varphi _{\lambda
}^{\rho }(k)\right\rangle $ is $\epsilon _{\lambda }^{\rho }=\rho \lbrack
\Upsilon _{k}+\lambda (\Upsilon _{k}^{2}-\kappa _{+}^{2}\kappa
_{-}^{2})^{1/2}-\gamma ^{2}]^{1/2}$ with $\Upsilon _{k}=(\left\vert \mu
_{k}\right\vert ^{2}+\kappa _{+}^{2}+\kappa _{-}^{2})/2$ and $\rho ,\lambda
=\pm $.

For nonzero $\kappa _{+}\kappa _{-}$ in Hermitian $H(k)$, this four-band
model can be regarded as two identical Rice-Mele models \cite{WRPRB}. The
topological features of the Rice-Mele model retain in the non-Hermitian
generalization. The Chern number $c_{\lambda }^{\rho }$ has a precise
physical meaning: the quantum particle transport for the energy band $(\rho
,\lambda )$ over an enclosed adiabatic passage along a closed cycle \cite%
{XiaoDRMP}. $c_{\lambda }^{\rho }$\ equals to the winding number of loop
around the band touching point $\left( \delta ,\kappa \right) =\left(
0,0\right) $ in the parameter space.

The biorthonormal current \cite{Book,KawabataPRB98} across sites $a_{N}$ and
$b_{N-1}$ is
\begin{equation}
j_{N}(t)=-i\sum_{m=1}^{N}\langle \eta _{m}(t)|\{J[1-\delta
(t)]a_{N}^{\dagger }b_{N-1}-\mathrm{H.c.}\}|\varphi _{m}(t)\rangle ,
\end{equation}%
where $m$ is the number of energy levels in the concerned energy band for
the $4N$ size system with periodic boundary condition. The parameters vary
as $\theta =\omega t$ in the numerical simulation under a quasi-adiabatic
process, where the speed of time evolution $\omega \ll 1$, and $t$ varies
from $0$ to a period of $T=2\pi \omega ^{-1}$. To demonstrate a
quasi-adiabatic process, we keep $f\left( t\right) =\left\vert \langle \bar{%
\eta}_{m}(t)\left\vert \varphi _{m}(t)\right\rangle \right\vert \rightarrow
1 $\ during the whole process by taking sufficient small $\omega $, where $%
\left\vert \bar{\eta}_{m}\left( t\right) \right\rangle $\ is the
corresponding instantaneous eigenstate of $\mathcal{H}^{\dagger }\left(
t\right) $. For the given initial eigenstates $|\varphi _{m}(0)\rangle
=|\varphi _{-,m}^{-}\rangle $ and $|\eta _{m}(0)\rangle =|\eta
_{-,m}^{-}\rangle $, the time evolved states are $\left\vert \varphi
_{m}(t)\right\rangle =$ $\mathcal{T}_{t}\exp [-i\int_{0}^{t}\mathcal{H}%
(t^{\prime })\mathrm{d}t^{\prime }]\left\vert \varphi _{m}(0)\right\rangle $
and $\left\vert \eta _{m}(t)\right\rangle =$ $\mathcal{T}_{t}\exp
[-i\int_{0}^{t}\mathcal{H}^{\dagger }(t^{\prime })\mathrm{d}t^{\prime
}]\left\vert \eta _{m}(0)\right\rangle $, where $\mathcal{T}_{t}$ is the
time ordering operator and $\mathcal{H}\left( t\right) $ is the Hamiltonian
in the real space. The accumulated charge pumping \cite{WRPRA,WRPRB} passing
the dimer $a_{N}b_{N-1}$\ during the interval $t$ is%
\begin{equation}
Q_{N}(t)=\int_{0}^{t}j_{N}(t^{\prime })dt^{\prime }.
\end{equation}

\begin{figure}[tb]
\includegraphics[bb=0 0 480 285, width=8.7 cm]{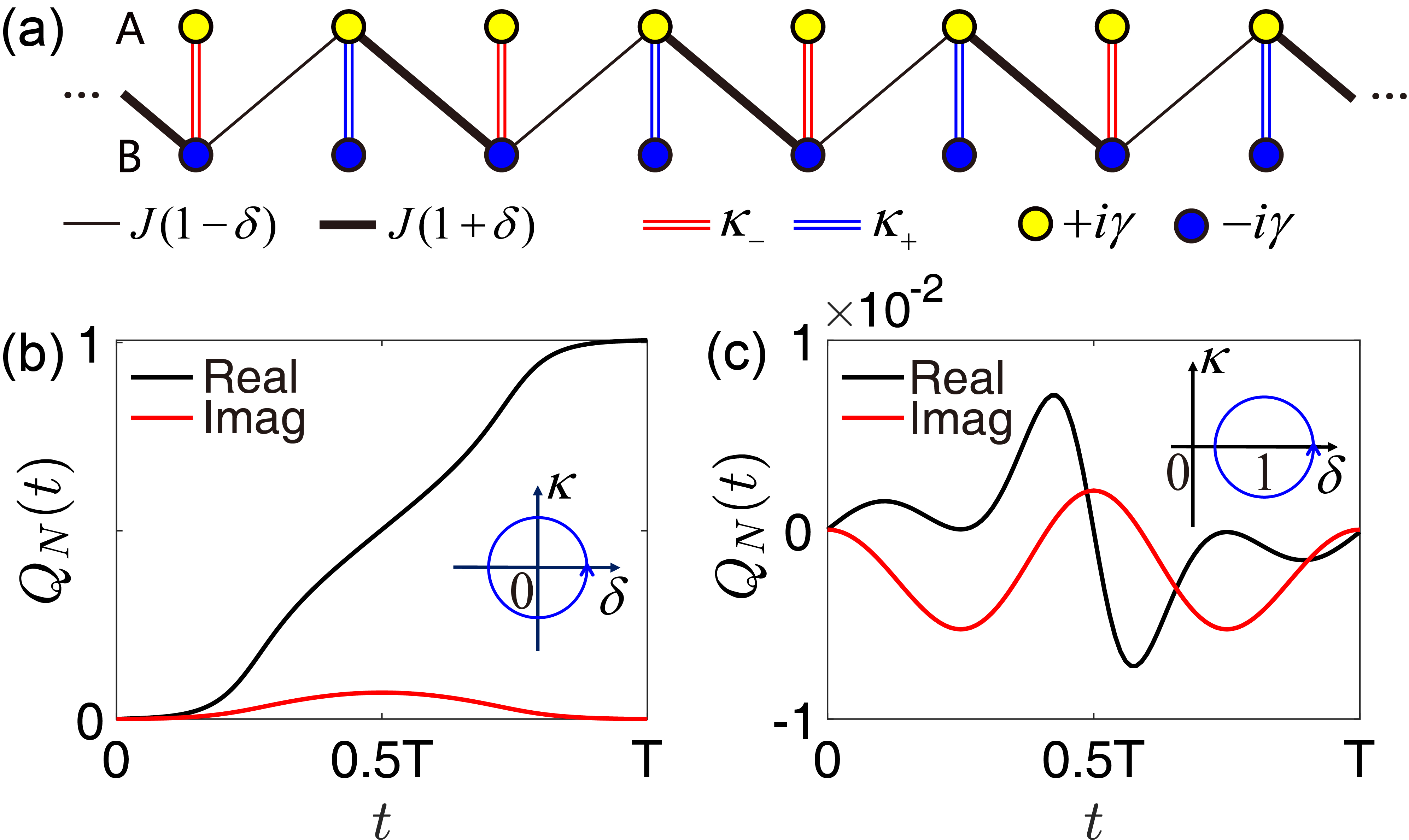}
\caption{(a) Schematic of the comb lattice. Numerical simulations of $Q_{N}(t)$ for the band $(\protect\rho, \protect\lambda) =(-,-)$ in two
quasi-adiabatic processes (insets) of (b) $\protect\delta_0=0$ and (c) $\protect\delta_0=1$ at $\protect\gamma=0.5$, $\protect\omega=10^{-3}$, and $T=2\protect\pi /\protect\omega$. Real (imaginary) part of $Q_N(t)$ is in
black (red); the corresponding $j_{N}(t)$ is shown in Appendix E.} \label%
{fig1}
\end{figure}

The topological charge pumping favorably agrees with the Chern number $%
c_{-}^{-}=1$ [Fig.~\ref{fig1}(b)] in the nontrivial phase or $c_{-}^{-}=0$
[Fig.~\ref{fig1}(c)] in the trivial phase for real energy band as that in
the Hermitian topological systems \cite{Thouless,Kraus,XDai}. For imaginary
energy band without EPs, the amplitude of evolved states $\left\vert \eta
_{m}(t)\right\rangle $ and $\left\vert \varphi _{m}(t)\right\rangle $
exponentially increase (or decrease); performing the quantization of
transport in a counterpart Hamiltonian $\mathcal{H}^{\prime }(k)=i\mathcal{H}%
(k)$ with corresponding real energy band is feasible to verify the Chern
number and the topological properties of imaginary energy band of $\mathcal{H%
}(k)$, since $\mathcal{H}^{\prime }(k)$ has identical topology and
eigenstate with $\mathcal{H}(k)$.

\begin{figure}[tb]
\centering
\includegraphics[bb=0 0 470 180, width=8.7 cm]{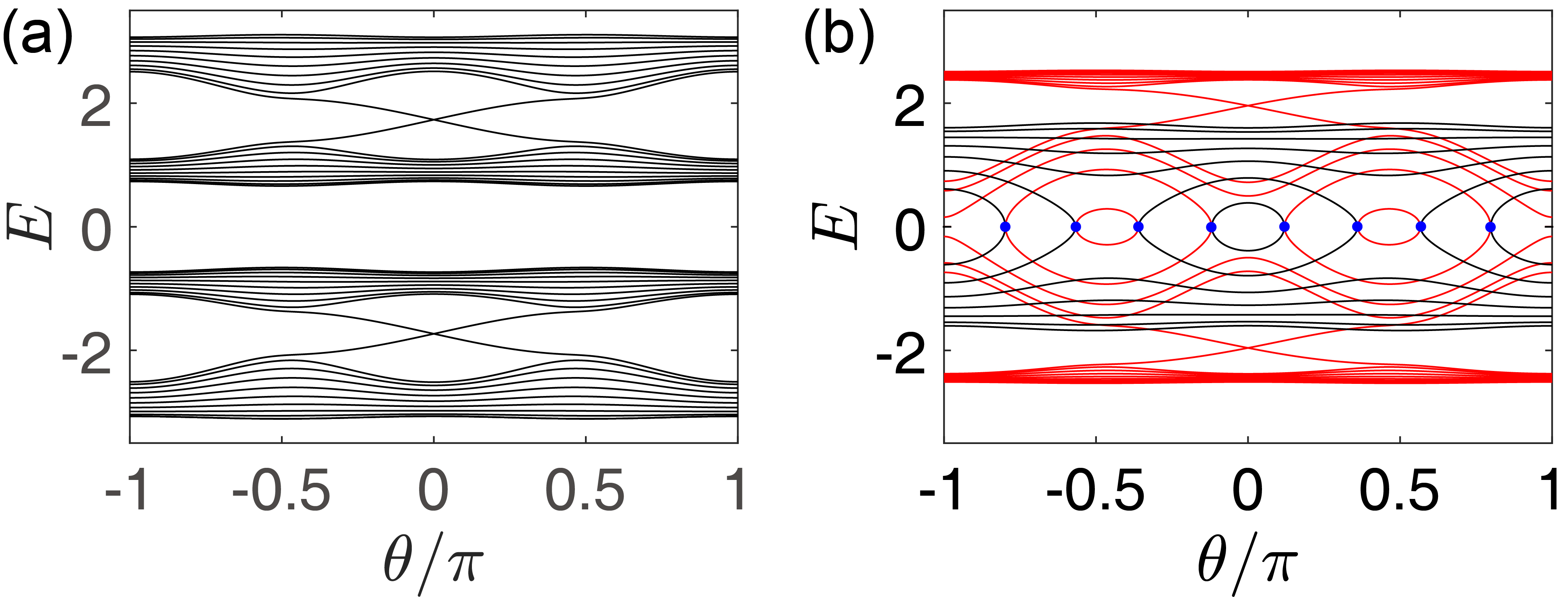}
\caption{Energy bands of the edge Hamiltonian for $\protect\delta_0=0$ at
(a) $\protect\gamma=1.0$ and (b) $\protect\gamma=2.8$, the real (imaginary)
part is in black (red), the blue dots are EPs. Other parameters are $J=1$, $\protect\kappa_0=2$, $R=0.6$, and $N=10$.}
\label{fig2}
\end{figure}

Alternatively, the topological charge pumping can be retrieved from the
dynamical evolution of edge states in the edge Hamiltonian $\mathcal{H}_{%
\mathrm{edge}}$ (see Appendix E), which is generated by truncating a
coupling $J(1+\delta )$ at the lattice boundary of the bulk Hamiltonian $%
\mathcal{H}$ in the real space [Fig.~\ref{fig1}(a)]. $\mathcal{H}_{\mathrm{%
edge}}$ and $H_{\mathrm{edge}}$ meet the condition of mapping since Eq.~(\ref%
{NH h}) still holds. Two pairs of edge states exist
\begin{eqnarray}
|\varphi _{\mathrm{R}}^{\pm }\rangle &=&\frac{1}{\sqrt{\Omega }}%
\sum\limits_{j=1}^{N}\varsigma ^{N-j}(e^{\pm i\vartheta _{\mathrm{R}%
}}a_{2j}^{\dagger }\pm b_{2j}^{\dagger })\left\vert \mathrm{vac}%
\right\rangle , \\
|\varphi _{\mathrm{L}}^{\pm }\rangle &=&\frac{1}{\sqrt{\Omega }}%
\sum\limits_{j=1}^{N}\varsigma ^{j-1}(e^{\pm i\vartheta _{\mathrm{L}%
}}a_{2j-1}^{\dagger }\pm b_{2j-1}^{\dagger })\left\vert \mathrm{vac}%
\right\rangle ,
\end{eqnarray}%
associated with the energies $\epsilon _{\mathrm{R}}^{\pm }=\pm (\kappa
_{+}^{2}-\gamma ^{2})^{1/2}$\ and $\epsilon _{\mathrm{L}}^{\pm }=\pm (\kappa
_{-}^{2}-\gamma ^{2})^{1/2}$, respectively; where $\varsigma =(\delta
-1)/(\delta +1)$, $\Omega =2(1-\varsigma ^{2N})/(1-\varsigma ^{2})$, $e^{\pm
i\vartheta _{\mathrm{R}}}=(\epsilon _{\mathrm{R}}^{+}\pm i\gamma )/\kappa
_{+}$ and $e^{\pm i\vartheta _{\mathrm{L}}}=(\epsilon _{\mathrm{L}}^{+}\pm
i\gamma )/\kappa _{-}$. The explicit expressions of edge states reveal a
fact that the mapping matrix only changes the local phase or amplitude. For
real $\epsilon _{\mathrm{R/L}}^{\pm }$, the edge state profiles are
independent of $\gamma $\ and $\kappa _{\pm }$ similar as that in Hermitian
\cite{XiaoDRMP} and non-Hermitian \cite{WRPRA} Rice-Mele models; for
imaginary $\epsilon _{\mathrm{R/L}}^{\pm }$, the probability becomes dense
in the sublattice with gain (loss) for $|\varphi _{\mathrm{R/L}}^{+}\rangle $
($|\varphi _{\mathrm{R/L}}^{-}\rangle $) \cite{BPeng}. The topological
charge pumping of an edge state for a loop $L$ in the $\kappa $-$\delta $
plane equals to the Chern number \cite{Hatsugai} (see Appendix E). The
energy bands are gapped and real at $\gamma =0$; as $\gamma $ increasing,
imaginary energy levels appear and non-Hermitian phase transition occurs. In
Figs.~\ref{fig2}(a) and~\ref{fig2}(b), the energy bands are depicted at weak
and strong $\gamma $, respectively. The not shown imaginary part for real
band is zero and vice versa. The edge states retain although energy bands
become imaginary. Recently, we notice an experimental work that reported the
existence of topological edge states in both unbroken and broken $\mathcal{PT%
}$-symmetric phases \cite{PXue1906}.

\textit{Discussion and conclusion.---}For chiral symmetric systems not in
the form of a bipartite lattice \cite{TonyPRL}, we can first apply a unitary
transformation to get the block off-diagonal form Hamiltonian [Eq.~(\ref{hq}%
)]; then, introduce the non-Hermiticity [Eq.~(\ref{NH h})]; after the
inversion unitary transformation, a non-Hermitian system possessing
identical topology to the chiral symmetric Hermitian system is generated.
Notably, the mapping theory is applicable for $H\mathbf{(q)}$ instead of the
core matrix $H\mathbf{(k)}$, where $\mathbf{q}$ is a set of periodic
parameters instead of the momentum $\mathbf{k}$. In addition to the gapped
topological systems, the non-Hermitian generalizations are applicable for
gapless topological systems \cite{MPan,OZ,Lieu,SLin17,ZKLSR,WPSR}. After
introducing the non-Hermiticity in the proposed manner, the gapless
degeneracy points may change into pairs of EPs, EP rings, or EP surfaces~%
\cite{Weimann,Szameit11,CerjanPRB,CerjanEPring}; although the non-Hermitian
phase transition occurs, the topology remains unchanged and can be
characterized by winding number as indicated in Refs. \cite%
{YXu,YXu19,CerjanPRB}.

Our findings provide insights into the interplay between non-Hermiticity and
topology. In contrast to the topological phase transition induced by the
non-Hermiticity, we propose the non-Hermitian generalization that completely
retains the topological phase transition and the (non)existence of edge
states in the original chiral symmetric Hermitian systems; and the
non-Hermitian phase transition does not alter or destroy the original
topology. This dramatically differs from the non-Hermiticity induced
topological phase transition~\cite{Takata,KawabataNC}, differs from the
breakdown of CBBC induced by gain and loss or non-Hermitian asymmetric
coupling \cite%
{TonyPRL,Martinez,KawabataPRB,HZhang,CHLee,LJL,ZGong,ZWang,WYi,Kunst}; and
differs from the situation that the strong non-Hermiticity destroys the
topological edge states due to the non-Hermitian phase transition associated
with the appearance of band touching EPs \cite{JLBBC}. The topological phase
transition and the non-Hermitian phase transition are independent and
separately controllable. This unique feature is valuable for the
explorations of novel non-Hermitian topological phases and topologically
protected edge state lasing.

\textit{Acknowledgment}.---This work was supported by National Natural
Science Foundation of China (Grants No.~11874225 and No.~11605094).

\section*{Appendix}

\renewcommand{\thesubsection}{\Alph{subsection}}

\subsection{Mapping matrix}

\setcounter{equation}{0} \renewcommand{\theequation}{A\arabic{equation}} %
\setcounter{figure}{0} \renewcommand{\thefigure}{A\arabic{figure}}

The Schr\"{o}dinger equation for the original Hermitian Hamiltonian $H(%
\mathbf{k})$ is $H(\mathbf{k})\left\vert \phi _{\lambda }^{\rho }(\mathbf{k}%
)\right\rangle =\varepsilon _{\lambda }^{\rho }(\mathbf{k})\left\vert \phi
_{\lambda }^{\rho }(\mathbf{k})\right\rangle $. $H(\mathbf{k})$ is the
parent Hamiltonian in the non-Hermitian generalization, and has the chiral
symmetry,%
\begin{equation}
H(\mathbf{k})=\left(
\begin{array}{cc}
0 & D(\mathbf{k}) \\
D^{\dagger }(\mathbf{k}) & 0%
\end{array}%
\right) ,
\end{equation}%
where we consider $D(\mathbf{k})$ as an arbitrary $n\times n$ matrix.

The eigenstates of $\mathcal{H}(\mathbf{k})=H(\mathbf{k})+i\gamma \left(
\sigma _{z}\otimes I_{n}\right) $ are
\begin{equation}
\left\vert \varphi _{\lambda }^{\rho }(\mathbf{k})\right\rangle =M_{\lambda
}^{\rho }\left( \mathbf{k,}\gamma \right) \left\vert \phi _{\lambda }^{\rho
}(\mathbf{k})\right\rangle ,
\end{equation}%
with eigenvalues
\begin{equation}
\epsilon _{\lambda }^{\rho }(\mathbf{k})=\rho \sqrt{\left[ \varepsilon
_{\lambda }^{\rho }\left( \mathbf{k}\right) \right] ^{2}-\gamma ^{2}},
\end{equation}%
where the mapping matrix has the form
\begin{equation}
M_{\lambda }^{\rho }\left( \mathbf{k,}\gamma \right) =\left(
\begin{array}{cc}
a_{\lambda }^{\rho }\left( \mathbf{k,}\gamma \right) I_{n} & 0 \\
0 & I_{n}%
\end{array}%
\right) ,
\end{equation}%
and $a_{\lambda }^{\rho }\left( \mathbf{k,}\gamma \right) =\left[ \epsilon
_{\lambda }^{\rho }\left( \mathbf{k}\right) +i\gamma \right] /\varepsilon
_{\lambda }^{\rho }\left( \mathbf{k}\right) $ fulfills $\left\vert
a_{\lambda }^{\rho }\left( \mathbf{k,}\gamma \right) \right\vert =1$ for
real $\epsilon _{\lambda }^{\rho }\left( \mathbf{k}\right) $ and is pure
imaginary for imaginary $\epsilon _{\lambda }^{\rho }\left( \mathbf{k}%
\right) $. For real $\epsilon _{\lambda }^{\rho }\left( \mathbf{k}\right) $,
the factor $a_{\lambda }^{\rho }\left( \mathbf{k,}\gamma \right) $ can be
written in the form of $a_{\lambda }^{\rho }\left( \mathbf{k,}\gamma \right)
=e^{i\vartheta }$, with $\vartheta =\arctan \left[ \gamma /\epsilon
_{\lambda }^{\rho }\left( \mathbf{k}\right) \right] $.

Notice that,%
\begin{equation}
H\left( \mathbf{k}\right) M_{\lambda }^{\rho }\left( \mathbf{k,}\gamma
\right) =\left(
\begin{array}{cc}
I_{n} & 0 \\
0 & a_{\lambda }^{\rho }\left( \mathbf{k,}\gamma \right) I_{n}%
\end{array}%
\right) H\left( \mathbf{k}\right) ,
\end{equation}%
we have $H\left( \mathbf{k}\right) \left\vert \varphi _{\lambda }^{\rho
}\left( \mathbf{k}\right) \right\rangle =\left[ H\left( \mathbf{k}\right)
M_{\lambda }^{\rho }\left( \mathbf{k,}\gamma \right) \right] \left\vert \phi
_{\lambda }^{\rho }\left( \mathbf{k}\right) \right\rangle $, then,%
\begin{eqnarray}
&&H\left( \mathbf{k}\right) \left\vert \varphi _{\lambda }^{\rho }\left(
\mathbf{k}\right) \right\rangle  \notag \\
&=&\varepsilon _{\lambda }^{\rho }\left( \mathbf{k}\right) \left(
\begin{array}{cc}
I_{n} & 0 \\
0 & a_{\lambda }^{\rho }\left( \mathbf{k,}\gamma \right) I_{n}%
\end{array}%
\right) \left\vert \phi _{\lambda }^{\rho }\left( \mathbf{k}\right)
\right\rangle \\
&=&\varepsilon _{\lambda }^{\rho }\left( \mathbf{k}\right) \left(
\begin{array}{cc}
\left[ a_{\lambda }^{\rho }\left( \mathbf{k,}\gamma \right) \right]
^{-1}I_{n} & 0 \\
0 & a_{\lambda }^{\rho }\left( \mathbf{k,}\gamma \right) I_{n}%
\end{array}%
\right) \left\vert \varphi _{\lambda }^{\rho }\left( \mathbf{k}\right)
\right\rangle ,  \notag
\end{eqnarray}%
therefore, we obtain%
\begin{widetext}
\begin{eqnarray}
\mathcal{H}\left( \mathbf{k}\right) \left\vert \varphi _{\lambda }^{\rho
}\left( \mathbf{k}\right) \right\rangle &=&\left[ H\left( \mathbf{k}\right)
+i\gamma \sigma _{z}\otimes I_{n}\right] \left\vert \varphi _{\lambda
}^{\rho }\left( \mathbf{k}\right) \right\rangle \notag\\&=&\left(
\begin{array}{cc}
\{\varepsilon _{\lambda }^{\rho }\left( \mathbf{k}\right) \left[ a_{\lambda
}^{\rho }\left( \mathbf{k,}\gamma \right) \right] ^{-1}+i\gamma \}I_{n} & 0
\\
0 & \left[ \varepsilon _{\lambda }^{\rho }\left( \mathbf{k}\right)
a_{\lambda }^{\rho }\left( \mathbf{k,}\gamma \right) -i\gamma \right] I_{n}%
\end{array}%
\right) \left\vert \varphi _{\lambda }^{\rho }\left( \mathbf{k}\right)
\right\rangle .
\end{eqnarray}%
\end{widetext}
From $a_{\lambda }^{\rho }\left( \mathbf{k,}\gamma \right) =\left[ \epsilon
_{\lambda }^{\rho }\left( \mathbf{k}\right) +i\gamma \right] /\varepsilon
_{\lambda }^{\rho }\left( \mathbf{k}\right) $ and $\epsilon _{\lambda
}^{\rho }(\mathbf{k})=\rho \sqrt{\left[ \varepsilon _{\lambda }^{\rho
}\left( \mathbf{k}\right) \right] ^{2}-\gamma ^{2}}$, we have $\varepsilon
_{\lambda }^{\rho }\left( \mathbf{k}\right) \left[ a_{\lambda }^{\rho
}\left( \mathbf{k,}\gamma \right) \right] ^{-1}+i\gamma =\varepsilon
_{\lambda }^{\rho }\left( \mathbf{k}\right) a_{\lambda }^{\rho }\left(
\mathbf{k,}\gamma \right) -i\gamma =\epsilon _{\lambda }^{\rho }\left(
\mathbf{k}\right) $. Thus,%
\begin{equation}
\mathcal{H}\left( \mathbf{k}\right) \left\vert \varphi _{\lambda }^{\rho
}\left( \mathbf{k}\right) \right\rangle =\epsilon _{\lambda }^{\rho }\left(
\mathbf{k}\right) \left\vert \varphi _{\lambda }^{\rho }\left( \mathbf{k}%
\right) \right\rangle .
\end{equation}%
In parallel, the eigenstate of $\mathcal{H}^{\dagger }\left( \mathbf{k}%
\right) $ is given by $\left\vert \eta _{\lambda }^{\rho }\left( \mathbf{k}%
\right) \right\rangle =\left[ M_{\lambda }^{\rho }\left( \mathbf{k,}\gamma
\right) \right] ^{\dagger }\left\vert \phi _{\lambda }^{\rho }\left( \mathbf{%
k}\right) \right\rangle $ with eigenvalue $\left[ \epsilon _{\lambda }^{\rho
}\left( \mathbf{k}\right) \right] ^{\ast }$.

\subsection{Unitary transformation}

\setcounter{equation}{0} \renewcommand{\theequation}{B\arabic{equation}} %
\setcounter{figure}{0} \renewcommand{\thefigure}{B\arabic{figure}}

\begin{figure*}[tbp]
\includegraphics[ bb=0 0 800 240, width=14.5 cm, clip]{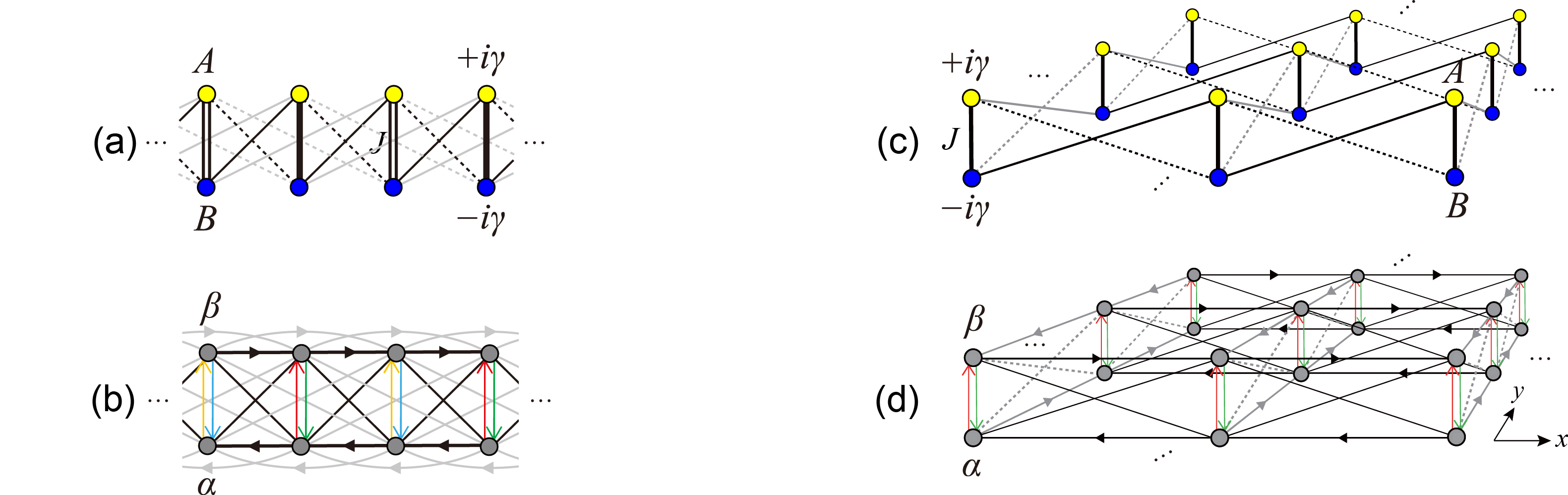}
\caption{(a) [(c)] Schematic of a bipartite lattice in 1D (2D), the
non-Hermitian gain and loss are introduced in the sublattices $A$ and $B$,
respectively. (b) [(d)] Schematic of the equivalent lattice of (a) [(c)],
the double arrows indicate the asymmetric couplings. For the sake of
clarity, the long range couplings in the 2D lattices are not shown in the
schematics.}
\label{figS1}
\end{figure*}

Figure \ref{figS1}(a) depicts a one-dimensional (1D) bipartite lattice. The
lines are the couplings between sublattices $A$ and $B$. Each pair of upper
and lower sites constitute a dimer. For a dimer with coupling $J$ and
balanced gain and loss $\pm i\gamma $, the dimer is $\mathcal{PT}$-symmetric
described by $J\sigma _{x}+i\gamma \sigma _{z}$. Applying a unitary
transformation
\begin{equation}
U=\left( I _{2}+i\sigma _{x}\right) /\sqrt{2},
\end{equation}%
we obtain a non-Hermitian dimer with asymmetric couplings $J\pm \gamma $ in
the form of
\begin{equation}
U\left( J\sigma _{x}+i\gamma \sigma _{z}\right) U^{-1}=J\sigma _{x}+i\gamma
\sigma _{y}.
\end{equation}%
Similarly, the lattice in Figure \ref{figS1}(a) changes into the lattice in
Figure \ref{figS1}(b) with asymmetric inter sublattice couplings. The gain
and loss change into asymmetric intra dimer couplings (vertical arrows); the
inter dimer couplings (slant lines) change to the Hermitian couplings
including the inter sublattice reciprocal cross-stitch couplings, and the
intra sublattice nonreciprocal couplings with symmetric amplitude
(horizontal arrows). The nonreciprocal couplings vanish if inter sublattice
couplings in the Hermitian system $H$ are symmetric (the situation that the
dashed and solid slant lines are identical). The imaginary gauge field is
created at $J\gamma \neq 0$ along the vertical direction, but not along the
horizontal direction (translational invariant direction); thus,
non-Hermitian AB effect is absent and the bulk-boundary correspondence is
valid. The conclusion is applicable in a general situation for systems with
nonreciprocal couplings and for higher dimensional bipartite lattices.

In a general case, the topological system may have complex coupling. For a
nonreciprocal coupling $Je^{\pm i\phi }$ with Peierls phase $e^{\pm i\phi }$
and coupling amplitude $J$, the $\mathcal{PT}$-symmetric dimer changes to
\begin{eqnarray}
&&U\left(
\begin{array}{cc}
i\gamma & Je^{-i\phi } \notag \\
Je^{i\phi } & -i\gamma%
\end{array}%
\right) U^{-1} \\
&=&\left(
\begin{array}{cc}
0 & e^{-i\phi }\left( J+\gamma \right) \\
e^{i\phi }\left( J-\gamma \right) & 0%
\end{array}%
\right) ,
\end{eqnarray}%
where the coupling with symmetric amplitude is changed into coupling with
asymmetric amplitude $J+\gamma $ and $J-\gamma $ associated with
nonreciprocal Peierls phase $e^{-i\phi }$ and $e^{i\phi }$, respectively.
The unitary transformation $U$ applied is%
\begin{equation}
U=\frac{1}{\sqrt{2}}\left(
\begin{array}{cc}
1 & ie^{-i\phi } \\
ie^{i\phi } & 1%
\end{array}%
\right) .
\end{equation}

In two-dimensional topological systems with chiral symmetry, for instance, a
two layer system with inter layer couplings as shown in Figure \ref{figS1}%
(c), which is a typical bipartite lattice. The non-Hermitian extension is to
introduce gain and loss in the upper and lower layers, respectively. Then,
the unitary transformation $U$ applied to each corresponding upper and lower
sites yields a new two layer lattice as shown in Figure \ref{figS1}(d), the
asymmetric couplings only exist between the new two layers (sublattices)
after unitary transformation. For higher dimensional systems, the asymmetric
couplings still only exist between the two new sublattices after the unitary
transformation, which is similar as the one-dimensional and two-dimensional
cases. Thus, the nonzero Aharonov-Bohm effect is absent in any translational
direction of the topological systems, and the conventional bulk-boundary
correspondence holds in the non-Hermitian generalization. The conclusion
coincides with that of the mapping theory.

\subsection{Mapping of geometric phase and Chern number}

\setcounter{equation}{0} \renewcommand{\theequation}{C\arabic{equation}} %
\setcounter{figure}{0} \renewcommand{\thefigure}{C\arabic{figure}}

In this section, we show that the Berry connection, Berry curvature and
Chern number of the non-Hermitian Hamiltonian in the momentum space $%
\mathcal{H}(\mathbf{k})=H(\mathbf{k})+i\gamma \left( \sigma _{z}\otimes
I_{n}\right) $ can be mapped from the Hermitian Hamiltonian $H(\mathbf{k})$
with chiral symmetry by using the mapping matrix. We prove that two
topological systems $H(\mathbf{k})$ and $\mathcal{H}(\mathbf{k})$ share an
identical Chern number, although their Berry connection and Berry curvature
are different. The conclusion is independent of the presence of exceptional
points (EPs) in the energy bands.

For the chiral symmetric system $H(\mathbf{k})$, we have%
\begin{equation}
\mathcal{S}H(\mathbf{k})\mathcal{S}^{-1}=-H(\mathbf{k}),
\end{equation}%
with $\mathcal{S}=\left( \sigma _{z}\otimes I_{n}\right) $. Then, from $H(%
\mathbf{k})\left\vert \phi _{\lambda }^{\rho }(\mathbf{k})\right\rangle
=\varepsilon _{\lambda }^{\rho }(\mathbf{k})\left\vert \phi _{\lambda
}^{\rho }(\mathbf{k})\right\rangle $, we have $\mathcal{S}H(\mathbf{k})%
\mathcal{S}^{-1}\mathcal{S}\left\vert \phi _{\lambda }^{\rho }(\mathbf{k}%
)\right\rangle =\varepsilon _{\lambda }^{\rho }(\mathbf{k})\mathcal{S}%
\left\vert \phi _{\lambda }^{\rho }(\mathbf{k})\right\rangle $, which gives $%
H(\mathbf{k})\left[ \mathcal{S}\left\vert \phi _{\lambda }^{\rho }(\mathbf{k}%
)\right\rangle \right] =-\varepsilon _{\lambda }^{\rho }(\mathbf{k})\left[
\mathcal{S}\left\vert \phi _{\lambda }^{\rho }(\mathbf{k})\right\rangle %
\right] $. Thus,
\begin{equation}
\left\vert \phi _{\lambda }^{-\rho }(\mathbf{k})\right\rangle =\mathcal{S}%
\left\vert \phi _{\lambda }^{\rho }(\mathbf{k})\right\rangle ,
\end{equation}%
is the eigenstate for energy $\varepsilon _{\lambda }^{-\rho }(\mathbf{k}%
)=-\varepsilon _{\lambda }^{\rho }(\mathbf{k})$.

We introduce the conventional Berry connection and Berry curvature, which
are called the \textrm{RR} Berry connection and Berry curvature \cite{Shen}.
The \textrm{RR} Berry connection for non-Hermitian system $\mathcal{H}(%
\mathbf{k})$ is
\begin{equation}
\left( \overrightarrow{\mathcal{A}}_{\lambda }^{\rho }\right) _{\mathrm{RR}%
}=i\left\langle \varphi _{\lambda }^{\rho }(\mathbf{k})\right\vert \nabla _{%
\mathbf{k}}\left\vert \varphi _{\lambda }^{\rho }(\mathbf{k})\right\rangle ,
\end{equation}%
in which
\begin{equation}
\left\vert \varphi _{\lambda }^{\rho }(\mathbf{k})\right\rangle =M_{\lambda
}^{\rho }(\mathbf{k},\gamma )\left\vert \phi _{\lambda }^{\rho }(\mathbf{k}%
)\right\rangle ,
\end{equation}%
and the normalization condition is satisfied, $\left\langle \varphi
_{\lambda }^{\rho }(\mathbf{k})\right. \left\vert \varphi _{\lambda }^{\rho
}(\mathbf{k})\right\rangle =1$ under the mapping matrix%
\begin{equation}
M_{\lambda }^{\rho }(\mathbf{k},\gamma )=\Pi _{\lambda }^{\rho }(\mathbf{k}%
,\gamma )\left(
\begin{array}{cc}
a_{\lambda }^{\rho }(\mathbf{k},\gamma )I_{n} & 0 \\
0 & I_{n}%
\end{array}%
\right) ,
\end{equation}%
with $\Pi _{\lambda }^{\rho }(\mathbf{k},\gamma )=1$ ($\Pi _{\lambda }^{\rho
}(\mathbf{k},\gamma )=\sqrt{2/\{1-\left[ a_{\lambda }^{\rho }(\mathbf{k}%
,\gamma )\right] ^{2}\}}$) for real (imaginary) $\epsilon _{\lambda }^{\rho
}(\mathbf{k})$ to guarantee the normalization condition. Then the \textrm{RR}
Berry connection can be written as%
\begin{eqnarray}
&&\left( \overrightarrow{\mathcal{A}}_{\lambda }^{\rho }\right) _{\mathrm{RR}%
}  \notag \\
&=&i\left\langle \phi _{\lambda }^{\rho }(\mathbf{k})\right\vert \left[
M_{\lambda }^{\rho }(\mathbf{k},\gamma )\right] ^{\dagger }\nabla _{\mathbf{k%
}}\left[ M_{\lambda }^{\rho }(\mathbf{k},\gamma )\left\vert \phi _{\lambda
}^{\rho }(\mathbf{k})\right\rangle \right]  \notag \\
&=&i\left\langle \phi _{\lambda }^{\rho }(\mathbf{k})\right\vert \left[
M_{\lambda }^{\rho }(\mathbf{k},\gamma )\right] ^{\dagger }M_{\lambda
}^{\rho }(\mathbf{k},\gamma )\nabla _{\mathbf{k}}\left\vert \phi _{\lambda
}^{\rho }(\mathbf{k})\right\rangle  \notag \\
&&+i\left\langle \phi _{\lambda }^{\rho }(\mathbf{k})\right\vert \left[
M_{\lambda }^{\rho }(\mathbf{k},\gamma )\right] ^{\dagger }\left[ \nabla _{%
\mathbf{k}}M_{\lambda }^{\rho }(\mathbf{k},\gamma )\right] \left\vert \phi
_{\lambda }^{\rho }(\mathbf{k})\right\rangle .
\end{eqnarray}

For real $\epsilon _{\lambda }^{\rho }(\mathbf{k})$, we have $a_{\lambda
}^{\rho }(\mathbf{k},\gamma )=e^{i\vartheta }$ with $\vartheta =\arctan %
\left[ \gamma /\epsilon _{\lambda }^{\rho }\left( \mathbf{k}\right) \right] $%
; then
\begin{eqnarray}
&&\left[ M_{\lambda }^{\rho }(\mathbf{k},\gamma )\right] ^{\dagger
}M_{\lambda }^{\rho }(\mathbf{k},\gamma )  \notag \\
&=&\left(
\begin{array}{cc}
e^{-i\vartheta }I_{n} & 0 \\
0 & I_{n}%
\end{array}%
\right) \left(
\begin{array}{cc}
e^{i\vartheta }I_{n} & 0 \\
0 & I_{n}%
\end{array}%
\right) =I_{2n},
\end{eqnarray}%
and
\begin{eqnarray}
&&\left[ M_{\lambda }^{\rho }(\mathbf{k},\gamma )\right] ^{\dagger }\nabla _{%
\mathbf{k}}M_{\lambda }^{\rho }(\mathbf{k},\gamma )  \notag \\
&=&\left(
\begin{array}{cc}
e^{-i\vartheta }I_{n} & 0 \\
0 & I_{n}%
\end{array}%
\right) \nabla _{\mathbf{k}}\left(
\begin{array}{cc}
e^{i\vartheta }I_{n} & 0 \\
0 & I_{n}%
\end{array}%
\right)  \notag \\
&=&\left(
\begin{array}{cc}
I_{n} & 0 \\
0 & 0%
\end{array}%
\right) i\nabla _{\mathbf{k}}\vartheta .
\end{eqnarray}%
Thus, the \textrm{RR} Berry connection is
\begin{equation}
\left( \overrightarrow{\mathcal{A}}_{\lambda }^{\rho }\right) _{\mathrm{RR}}=%
\mathbf{A}_{\lambda }^{\rho }-\frac{1}{2}\nabla _{\mathbf{k}}\vartheta ,
\end{equation}%
with $\mathbf{A}_{\lambda }^{\rho }=i\left\langle \phi _{\lambda }^{\rho }(%
\mathbf{k})\right\vert \nabla _{\mathbf{k}}\left\vert \phi _{\lambda }^{\rho
}(\mathbf{k})\right\rangle $ being the Berry connection of the Hermitian
system $H\left( \mathbf{k}\right) $.

For imaginary $\epsilon _{\lambda }^{\rho }(\mathbf{k})$, we have $%
[a_{\lambda }^{\rho }\left( \mathbf{k,}\gamma \right) ]^{\ast }=-a_{\lambda
}^{\rho }\left( \mathbf{k,}\gamma \right) $,%
\begin{eqnarray}
&&\left[ M_{\lambda }^{\rho }(\mathbf{k},\gamma )\right] ^{\dagger
}M_{\lambda }^{\rho }(\mathbf{k},\gamma )i\nabla _{\mathbf{k}}-I_{2n}i\nabla
_{\mathbf{k}}  \notag \\
&=&\frac{1+\left[ a_{\lambda }^{\rho }(\mathbf{k},\gamma )\right] ^{2}}{%
\left[ a_{\lambda }^{\rho }(\mathbf{k},\gamma )\right] ^{2}-1}\left(
\begin{array}{cc}
I_{n} & 0 \\
0 & -I_{n}%
\end{array}%
\right) i\nabla _{\mathbf{k}}  \notag \\
&=&\frac{\epsilon _{\lambda }^{\rho }(\mathbf{k})}{i\gamma }\mathcal{S}%
i\nabla _{\mathbf{k}},
\end{eqnarray}%
and
\begin{widetext}
\begin{eqnarray}
\left[ M_{\lambda }^{\rho }(\gamma ,\mathbf{k})\right] ^{\dagger }\left[
\nabla _{\mathbf{k}}M_{\lambda }^{\rho }(\gamma ,\mathbf{k})\right]  &=&%
\left[ \Pi _{\lambda }^{\rho }(\mathbf{k},\gamma )\right] ^{\ast }\left(
\begin{array}{cc}
\left[ a_{\lambda }^{\rho }(\mathbf{k},\gamma )\right] ^{\ast }I_{n} & 0 \\
0 & I_{n}%
\end{array}%
\right) \nabla _{\mathbf{k}}\left(
\begin{array}{cc}
a_{\lambda }^{\rho }(\mathbf{k},\gamma )\Pi _{\lambda }^{\rho }(\mathbf{k}%
,\gamma )I_{n} & 0 \\
0 & \Pi _{\lambda }^{\rho }(\mathbf{k},\gamma )I_{n}%
\end{array}%
\right)   \notag \\
&=&-\frac{1}{2}a_{\lambda }^{\rho }(\mathbf{k},\gamma )\left[ \Pi _{\lambda
}^{\rho }(\mathbf{k},\gamma )\right] ^{4}\left[ \nabla _{\mathbf{k}%
}a_{\lambda }^{\rho }(\mathbf{k},\gamma )\right] \mathcal{S}.
\end{eqnarray}%
Then, the \textrm{RR} Berry connection
\begin{eqnarray}
\left( \overrightarrow{\mathcal{A}}_{\lambda }^{\rho }\right) _{\mathrm{RR}}
&=&i\left\langle \phi _{\lambda }^{\rho }(\mathbf{k})\right\vert \nabla _{%
\mathbf{k}}\left\vert \phi _{\lambda }^{\rho }(\mathbf{k})\right\rangle +%
\frac{\epsilon _{\lambda }^{\rho }(\mathbf{k})}{i\gamma }\left\langle \phi
_{\lambda }^{-\rho }(\mathbf{k})\right\vert \nabla _{\mathbf{k}}\left\vert
\phi _{\lambda }^{\rho }(\mathbf{k})\right\rangle -\frac{1}{2}ia_{\lambda
}^{\rho }(\mathbf{k},\gamma )\left[ \Pi _{\lambda }^{\rho }(\mathbf{k}%
,\gamma )\right] ^{4}\left[ \nabla _{\mathbf{k}}a_{\lambda }^{\rho }(\mathbf{%
\ k},\gamma )\right] \left\langle \phi _{\lambda }^{\rho }(\mathbf{k}%
)\right. \left\vert \phi _{\lambda }^{-\rho }(\mathbf{k})\right\rangle
\notag \\
&=&\mathbf{A}_{\lambda }^{\rho }+\frac{\epsilon _{\lambda }^{\rho }(\mathbf{k%
})}{i\gamma }\left\langle \phi _{\lambda }^{-\rho }(\mathbf{k})\right\vert
\nabla _{\mathbf{k}}\left\vert \phi _{\lambda }^{\rho }(\mathbf{k}%
)\right\rangle .
\end{eqnarray}
\end{widetext}
The definition of \textrm{RR} Berry connection is independent of the
biorthonormal basis. Although the eigenstates $\left\vert \varphi _{\lambda
}^{\rho }(\mathbf{k})\right\rangle ^{\mathrm{EP}}$ and $\left\vert \varphi
_{\lambda }^{-\rho }(\mathbf{k})\right\rangle ^{\mathrm{EP}}$ coalesce at
EPs and the biorthonormal basis is absent, the \textrm{RR} Berry connection
can still be defined. At EPs, the energy is $\epsilon _{\lambda }^{\rho }(%
\mathbf{k})^{\mathrm{EP}}=0$, and the mapping matrix has a simple form
\begin{equation}
\left[ M_{\lambda }^{\rho }(\mathbf{k},\gamma )\right] ^{\mathrm{EP}}=\left(
\begin{array}{cc}
i\rho I_{n} & 0 \\
0 & I_{n}%
\end{array}%
\right) .
\end{equation}%
Direct derivation shows that the \textrm{RR} Berry connection at EPs is $%
\left( \overrightarrow{\mathcal{A}}_{\lambda }^{\rho }\right) _{\mathrm{RR}%
}^{\mathrm{EP}}=i\left\langle \phi _{\lambda }^{\rho }(\mathbf{k}%
)\right\vert \nabla _{\mathbf{k}}\left\vert \phi _{\lambda }^{\rho }(\mathbf{%
\ k})\right\rangle =\mathbf{A}_{\lambda }^{\rho }$.

In conclusion, we have
\begin{equation}
\left( \overrightarrow{\mathcal{A}}_{\lambda }^{\rho }\right) _{\mathrm{RR}%
}=\left\{
\begin{array}{ll}
\mathbf{A}_{\lambda }^{\rho }-\frac{1}{2}\nabla _{\mathbf{k}}\vartheta , &
\text{real }\epsilon _{\lambda }^{\rho }(\mathbf{k}) \\
\mathbf{A}_{\lambda }^{\rho }+\left[ \mathbf{N}_{\lambda }^{\rho }(\mathbf{k}%
)\right] _{\mathrm{RR}}, & \text{imaginary }\epsilon _{\lambda }^{\rho }(%
\mathbf{k}) \\
\mathbf{A}_{\lambda }^{\rho }, & \text{at }\mathrm{EPs}%
\end{array}%
\right. ,
\end{equation}%
where $\left[ \mathbf{N}_{\lambda }^{\rho }(\mathbf{k})\right] _{\mathrm{RR}%
}=\epsilon _{\lambda }^{\rho }(\mathbf{k})\left\langle \phi _{\lambda
}^{-\rho }(\mathbf{k})\right\vert \nabla _{\mathbf{k}}\left\vert \phi
_{\lambda }^{\rho }(\mathbf{k})\right\rangle /(i\gamma )$. The \textrm{RR}
Berry connection $\left( \overrightarrow{\mathcal{A}}_{\lambda }^{\rho
}\right) _{\mathrm{RR}}$ is gauge dependent. If we take the gauge
transformation $\left\vert \varphi _{\lambda }^{\rho }(\mathbf{k}%
)\right\rangle \rightarrow e^{i\chi \left( \mathbf{k}\right) }\left\vert
\varphi _{\lambda }^{\rho }(\mathbf{k})\right\rangle $ with real $\chi
\left( \mathbf{k}\right) $, then we have an additional term $i\nabla _{
\mathbf{k}}\chi \left( \mathbf{k}\right) $ in $\left( \overrightarrow{
\mathcal{A}}_{\lambda }^{\rho }\right) _{\mathrm{RR}}$. However, $\left(
\overrightarrow{\mathcal{B}}_{\lambda }^{\rho }\right) _{\mathrm{RR}}=\nabla
_{\mathbf{k}}\times \left( \overrightarrow{\mathcal{A}}_{\lambda }^{\rho
}\right) _{\mathrm{RR}}$ is gauge independent.

For real $\epsilon _{\lambda }^{\rho }(\mathbf{k})$, we have $\left(
\overrightarrow{\mathcal{B}}_{\lambda }^{\rho }\right) _{\mathrm{RR}}=\nabla
_{\mathbf{k}}\times \left( \overrightarrow{\mathcal{A}}_{\lambda }^{\rho
}\right) _{\mathrm{RR}}=\nabla _{\mathbf{k}}\times \left( \mathbf{A}%
_{\lambda }^{\rho }-\frac{1}{2}\nabla _{\mathbf{k}}\vartheta \right) =\nabla
_{\mathbf{k}}\times \mathbf{A}_{\lambda }^{\rho }=B_{\lambda }^{\rho }$. For
imaginary $\epsilon _{\lambda }^{\rho }(\mathbf{k})$, the additional term $%
\nabla _{\mathbf{k}}\times \left[ \mathbf{N}_{\lambda }^{\rho }(\mathbf{k})%
\right] _{\mathrm{RR}}$ yields zero integration over the Brillouin zone. We
prove this as follows. Notice that the eigenstates $\left\vert \phi
_{\lambda }^{\rho }(\mathbf{k})\right\rangle $ and $\left\vert \phi
_{\lambda }^{-\rho }(\mathbf{k})\right\rangle $ can choose an identical
gauge in the same region of the Brillouin zone; it is because that the two
eigenstates are related through the chiral operator $\mathcal{S}$, and they
are orthogonal $\left\langle \phi _{\lambda }^{-\rho }(\mathbf{k})\right.
\left\vert \phi _{\lambda }^{\rho }(\mathbf{k})\right\rangle =0$. If we take
a gauge transformation%
\begin{equation}
\left\vert \phi _{\lambda }^{\rho }(\mathbf{k})^{\mathrm{I}}\right\rangle
=e^{i\chi \left( \mathbf{k}\right) }\left\vert \phi _{\lambda }^{\rho }(%
\mathbf{k})^{\mathrm{II}}\right\rangle ,
\end{equation}%
then, we have%
\begin{eqnarray}
\left\vert \phi _{\lambda }^{-\rho }(\mathbf{k})^{\mathrm{I}}\right\rangle
&=&\mathcal{S}\left\vert \phi _{\lambda }^{\rho }(\mathbf{k})^{\mathrm{I}%
}\right\rangle =\mathcal{S}e^{i\chi \left( \mathbf{k}\right) }\left\vert
\phi _{\lambda }^{\rho }(\mathbf{k})^{\mathrm{II}}\right\rangle \\
&=&e^{i\chi \left( \mathbf{k}\right) }\mathcal{S}\left\vert \phi _{\lambda
}^{\rho }(\mathbf{k})^{\mathrm{II}}\right\rangle =e^{i\chi \left( \mathbf{k}%
\right) }\left\vert \phi _{\lambda }^{-\rho }(\mathbf{k})^{\mathrm{II}%
}\right\rangle .  \notag
\end{eqnarray}%
Unlike the Berry connection, term $\left[ \mathbf{N}_{\lambda }^{\rho }(%
\mathbf{k})\right] _{\mathrm{RR}}$ is gauge independent
\begin{eqnarray}
&&\left[ N_{\lambda }^{\rho }(\mathbf{k})^{\mathrm{I}}\right] _{\mathrm{RR}}
\notag \\
&=&\epsilon _{\lambda }^{\rho }(\mathbf{k})\left\langle \phi _{\lambda
}^{-\rho }(\mathbf{k})^{\mathrm{I}}\right\vert i\nabla _{\mathbf{k}%
}\left\vert \phi _{\lambda }^{\rho }(\mathbf{k})^{\mathrm{I}}\right\rangle
/\left( i\gamma \right)  \notag \\
&=&\epsilon _{\lambda }^{\rho }(\mathbf{k})e^{-i\chi \left( \mathbf{k}%
\right) }\left\langle \phi _{\lambda }^{-\rho }(\mathbf{k})^{\mathrm{II}%
}\right\vert i\nabla _{\mathbf{k}}\left[ e^{i\chi \left( \mathbf{k}\right)
}\left\vert \phi _{\lambda }^{\rho }(\mathbf{k})^{\mathrm{II}}\right\rangle %
\right] /\left( i\gamma \right)  \notag \\
&=&\epsilon _{\lambda }^{\rho }(\mathbf{k})e^{-i\chi \left( \mathbf{k}%
\right) }\left[ i\nabla _{\mathbf{k}}e^{i\chi \left( \mathbf{k}\right) }%
\right] \left\langle \phi _{\lambda }^{-\rho }(\mathbf{k})^{\mathrm{II}%
}\right. \left\vert \phi _{\lambda }^{\rho }(\mathbf{k})^{\mathrm{II}%
}\right\rangle /\left( i\gamma \right)  \notag \\
&&+\epsilon _{\lambda }^{\rho }(\mathbf{\ k})\left\langle \phi _{\lambda
}^{-\rho }(\mathbf{k})^{\mathrm{II}}\right\vert i\nabla _{\mathbf{k}%
}\left\vert \phi _{\lambda }^{\rho }(\mathbf{k})^{\mathrm{II}}\right\rangle
/\left( i\gamma \right)  \notag \\
&=&\left[ N_{\lambda }^{\rho }(\mathbf{k})^{\mathrm{II}}\right] _{\mathrm{RR}%
}.
\end{eqnarray}%
One can use different gauges to define the eigenstate if the eigenstate
under one gauge is not well-defined in certain regions of the Brillouin
zone. We consider a case that the eigenstate of the concerned energy band is
well-defined under gauge $\mathrm{I}$ in the region $D^{\mathrm{I}}$ and
under gauge $\mathrm{II}$ in the rest region $D^{\mathrm{II}}$ (the cases
with more than two gauges required can be similarly generalized). Applying
Stokes theorem, we have%
\begin{eqnarray}
&&\oint\nolimits_{\mathrm{BZ}}\nabla _{\mathbf{k}}\times \left[ \mathbf{N}%
_{\lambda }^{\rho }(\mathbf{k})\right] _{\mathrm{RR}}d^{2}\mathbf{k}  \notag
\\
&=&\oint\nolimits_{\partial D^{\mathrm{I}}}\left[ \mathbf{N}_{\lambda
}^{\rho }(\mathbf{k})\right] _{\mathrm{RR}}^{\mathrm{I}}d\mathbf{k+}%
\oint\nolimits_{\partial D^{\mathrm{II}}}\left[ \mathbf{N}_{\lambda }^{\rho
}(\mathbf{k})\right] _{\mathrm{RR}}^{\mathrm{II}}d\mathbf{k}  \notag \\
&=&\oint\nolimits_{\partial D^{\mathrm{I}}}\left\{ \left[ \mathbf{N}%
_{\lambda }^{\rho }(\mathbf{k})\right] _{\mathrm{RR}}^{\mathrm{I}}-\left[
\mathbf{N}_{\lambda }^{\rho }(\mathbf{k})\right] _{\mathrm{RR}}^{\mathrm{II}%
}\right\} d\mathbf{k}=0.
\end{eqnarray}

The chiral symmetry of the Hermitian system $H(\mathbf{k})$ plays a crucial
role to obtain the above conclusions, the chiral symmetry makes term $\left[
\mathbf{N}_{\lambda }^{\rho }(\mathbf{k})\right] _{\mathrm{RR}}$ gauge
independent, thus, $\left[ \mathbf{N}_{\lambda }^{\rho }(\mathbf{k})\right]
_{\mathrm{RR}}$ does not contribute to the Chern number. Based on the above
analysis, the \textrm{RR} Chern number of the non-Hermitian system $\mathcal{%
\ H}(\mathbf{k})$ is \textit{exactly} identical to the Chern number of the
Hermitian system $H(\mathbf{k})$, even though there exists EPs and the
energy bands are inseparable (the energy bands of the corresponding
Hermitian system are separable, and the Chern number is well defined)
\begin{eqnarray}
\left( c_{\lambda }^{\rho }\right) _{\mathrm{RR}} &=&\frac{1}{2\pi }%
\oint\nolimits_{\mathrm{BZ}}\left( \overrightarrow{\mathcal{B}}_{\lambda
}^{\rho }\right) _{\mathrm{RR}}d^{2}\mathbf{k}  \notag \\
&=&\frac{1}{2\pi }\oint\nolimits_{\mathrm{BZ}}\mathbf{B}_{\lambda }^{\rho
}d^{2}\mathbf{k}.
\end{eqnarray}

Now, we discuss the \textrm{LR} Berry connection and Berry curvature based
on the biorthonormal basis \cite{KawabataPRB98}. $\left\vert \varphi
_{\lambda }^{\rho }(\mathbf{k})\right\rangle $ and $\left\vert \eta
_{\lambda }^{\rho }(\mathbf{k})\right\rangle $ are the eigenstates of $%
\mathcal{H}(\mathbf{k})$ and $\mathcal{H}^{\dagger }(\mathbf{k})$ with
eigenvalues $\epsilon _{\lambda }^{\rho }(\mathbf{k})$ and $\left[ \epsilon
_{\lambda }^{\rho }(\mathbf{k})\right] ^{\ast }$, respectively. The Schr\"{o}%
dinger equations are%
\begin{eqnarray}
\mathcal{H}\left( \mathbf{k}\right) \left\vert \varphi _{\lambda }^{\rho }(%
\mathbf{k})\right\rangle &=&\epsilon _{\lambda }^{\rho }(\mathbf{k}%
)\left\vert \varphi _{\lambda }^{\rho }(\mathbf{k})\right\rangle , \\
\mathcal{H}^{\dagger }\left( \mathbf{k}\right) \left\vert \eta _{\lambda
}^{\rho }(\mathbf{k})\right\rangle &=&\left[ \epsilon _{\lambda }^{\rho }(%
\mathbf{k})\right] ^{\ast }\left\vert \eta _{\lambda }^{\rho }(\mathbf{k}%
)\right\rangle .
\end{eqnarray}%
The eigenstates $\left\{ \left\vert \varphi _{\lambda }^{\rho }(\mathbf{k}%
)\right\rangle ,\left\vert \eta _{\lambda }^{\rho }(\mathbf{k})\right\rangle
\right\} $ can be mapped from the eigenstates of the original Hermitian
system $H\left( \mathbf{k}\right) $,
\begin{eqnarray}
\left\vert \varphi _{\lambda }^{\rho }(\mathbf{k})\right\rangle
&=&M_{\lambda }^{\rho }(\mathbf{k},\gamma )\left\vert \phi _{\lambda }^{\rho
}(\mathbf{k})\right\rangle , \\
\left\vert \eta _{\lambda }^{\rho }(\mathbf{k})\right\rangle &=&\left[
M_{\lambda }^{\rho }(\mathbf{k},\gamma )\right] ^{\dagger }\left\vert \phi
_{\lambda }^{\rho }(\mathbf{k})\right\rangle .
\end{eqnarray}%
In the absence of EP, the mapping matrix can be written as
\begin{equation}
M_{\lambda }^{\rho }(\mathbf{k},\gamma )=\Lambda _{\lambda }^{\rho }(\mathbf{%
\ k},\gamma )\left(
\begin{array}{cc}
a_{\lambda }^{\rho }(\mathbf{k},\gamma )I_{n} & 0 \\
0 & I_{n}%
\end{array}%
\right) ,
\end{equation}%
with $\Lambda _{\lambda }^{\rho }(\mathbf{k},\gamma )=\sqrt{2/\{1+\left[
a_{\lambda }^{\rho }(\mathbf{k},\gamma )\right] ^{2}\}}$ to guarantee the
biorthonormal normalization.

Based on the orthonormal relation $\langle \phi _{\lambda }^{\rho }(\mathbf{k%
})|\phi _{\lambda ^{\prime }}^{\rho ^{\prime }}(\mathbf{k}^{\prime })\rangle
=\delta _{\mathbf{kk}^{\prime }}\delta _{\lambda \lambda ^{\prime }}\delta
_{\rho \rho ^{\prime }}$ for the eigenstates of the original Hermitian
system, we have the biorthonormal relation%
\begin{equation}
\langle \eta _{\lambda }^{\rho }(\mathbf{k})|\varphi _{\lambda ^{\prime
}}^{\rho ^{\prime }}(\mathbf{k}^{\prime })\rangle =\delta _{\mathbf{kk}
^{\prime }}\delta _{\lambda \lambda ^{\prime }}\delta _{\rho \rho ^{\prime
}},
\end{equation}%
for the left and right eigenstates of the non-Hermitian system.

The Berry connection based on the left and right eigenstates is defined as
\begin{eqnarray}
\left( \overrightarrow{\mathcal{A}}_{\lambda }^{\rho }\right) _{\mathrm{LR}}
&=&i\left\langle \eta _{\lambda }^{\rho }(\mathbf{k})\right\vert \nabla _{%
\mathbf{k}}\left\vert \varphi _{\lambda }^{\rho }(\mathbf{k})\right\rangle \\
&=&i\left\langle \phi _{\lambda }^{\rho }(\mathbf{k})\right\vert M_{\lambda
}^{\rho }(\mathbf{k},\gamma )\nabla _{\mathbf{k}}\left[ M_{\lambda }^{\rho }(%
\mathbf{k},\gamma )\left\vert \phi _{\lambda }^{\rho }(\mathbf{k}%
)\right\rangle \right]  \notag \\
&=&i\left\langle \phi _{\lambda }^{\rho }(\mathbf{k})\right\vert \left[
M_{\lambda }^{\rho }(\mathbf{k},\gamma )\right] ^{2}\nabla _{\mathbf{k}%
}\left\vert \phi _{\lambda }^{\rho }(\mathbf{k})\right\rangle  \notag \\
&&+i\left\langle \phi _{\lambda }^{\rho }(\mathbf{k})\right\vert M_{\lambda
}^{\rho }(\mathbf{\ k},\gamma )\left[ \nabla _{\mathbf{k}}M_{\lambda }^{\rho
}(\mathbf{k},\gamma )\right] \left\vert \phi _{\lambda }^{\rho }(\mathbf{k}%
)\right\rangle ,  \notag
\end{eqnarray}%
where%
\begin{widetext}
\begin{eqnarray}
M_{\lambda }^{\rho }(\mathbf{k},\gamma )\left[ \nabla _{\mathbf{k}%
}M_{\lambda }^{\rho }(\mathbf{k},\gamma )\right]  &=&\Lambda _{\lambda
}^{\rho }(\mathbf{k},\gamma )\left(
\begin{array}{cc}
a_{\lambda }^{\rho }(\mathbf{k},\gamma )I_{n} & 0 \\
0 & I_{n}%
\end{array}%
\right) \nabla _{\mathbf{k}}\left(
\begin{array}{cc}
a_{\lambda }^{\rho }(\mathbf{k},\gamma )\Lambda _{\lambda }^{\rho }(\mathbf{k%
},\gamma )I_{n} & 0 \\
0 & \Lambda _{\lambda }^{\rho }(\mathbf{k},\gamma )I_{n}%
\end{array}%
\right)   \notag \\
&=&\frac{1}{2}a_{\lambda }^{\rho }(\mathbf{k},\gamma )\left[ \Lambda
_{\lambda }^{\rho }(\mathbf{k},\gamma )\right] ^{4}\left[ \nabla _{\mathbf{k}%
}a_{\lambda }^{\rho }(\mathbf{k},\gamma )\right] \mathcal{S},
\end{eqnarray}%
\end{widetext}
and%
\begin{eqnarray}
&&\left[ M_{\lambda }^{\rho }(\mathbf{k},\gamma )\right] ^{2}i\nabla _{%
\mathbf{\ k}}-I_{2n}i\nabla _{\mathbf{k}}  \notag \\
&=&\frac{\left[ a_{\lambda }^{\rho }(\mathbf{k},\gamma )\right] ^{2}-1}{1+%
\left[ a_{\lambda }^{\rho }(\mathbf{k},\gamma )\right] ^{2}}\left(
\begin{array}{cc}
I_{n} & 0 \\
0 & -I_{n}%
\end{array}%
\right) i\nabla _{\mathbf{k}}  \notag \\
&=&\frac{i\gamma }{\epsilon _{\lambda }^{\rho }(\mathbf{k})}\mathcal{S}%
i\nabla _{\mathbf{k}}.
\end{eqnarray}

Thus, the Berry connection is reduced to%
\begin{eqnarray}
\left( \overrightarrow{\mathcal{A}}_{\lambda }^{\rho }\right) _{\mathrm{LR}}
&=&i\left\langle \phi _{\lambda }^{\rho }(\mathbf{k})\right\vert \left[
M_{\lambda }^{\rho }(\mathbf{k},\gamma )\right] ^{2}\nabla _{\mathbf{k}%
}\left\vert \phi _{\lambda }^{\rho }(\mathbf{k})\right\rangle  \notag \\
&+&i\left\langle \phi _{\lambda }^{\rho }(\mathbf{k})\right\vert M_{\lambda
}^{\rho }(\mathbf{\ k},\gamma )\left[ \nabla _{\mathbf{k}}M_{\lambda }^{\rho
}(\mathbf{k},\gamma )\right] \left\vert \phi _{\lambda }^{\rho }(\mathbf{k}%
)\right\rangle  \notag \\
&=&\mathbf{A}_{\lambda }^{\rho }+\frac{i\gamma }{\epsilon _{\lambda }^{\rho
}(\mathbf{k})}\left\langle \phi _{\lambda }^{-\rho }(\mathbf{k})\right\vert
i\nabla _{\mathbf{k}}\left\vert \phi _{\lambda }^{\rho }(\mathbf{k}%
)\right\rangle  \notag \\
&+&\frac{1}{2}ia_{\lambda }^{\rho }\left[ \Lambda _{\lambda }^{\rho }(%
\mathbf{k},\gamma )\right] ^{4}\left[ \nabla _{\mathbf{k}}a_{\lambda }^{\rho
}(\mathbf{k},\gamma )\right] \left\langle \phi _{\lambda }^{-\rho }(\mathbf{k%
})\right. \left\vert \phi _{\lambda }^{\rho }(\mathbf{k})\right\rangle
\notag \\
&=&\mathbf{A}_{\lambda }^{\rho }+\left[ \mathbf{N}_{\lambda }^{\rho }(%
\mathbf{k})\right] _{\mathrm{LR}},
\end{eqnarray}%
where $\mathbf{A}_{\lambda }^{\rho }=i\left\langle \phi _{\lambda }^{\rho }(%
\mathbf{k})\right\vert \nabla _{\mathbf{k}}\left\vert \phi _{\lambda }^{\rho
}(\mathbf{k})\right\rangle $ is the Berry connection for the original
Hamiltonian $H(\mathbf{k})$, and
\begin{equation}
\left[ \mathbf{N}_{\lambda }^{\rho }(\mathbf{k})\right] _{\mathrm{LR}%
}=i\gamma \left\langle \phi _{\lambda }^{-\rho }(\mathbf{k})\right\vert
i\nabla _{\mathbf{k}}\left\vert \phi _{\lambda }^{\rho }(\mathbf{k}%
)\right\rangle /\epsilon _{\lambda }^{\rho }(\mathbf{k}).
\end{equation}%
The \textrm{LR} Berry connection $\left( \overrightarrow{\mathcal{A}}%
_{\lambda }^{\rho }\right) _{\mathrm{LR}}$ is gauge dependent. For instance,
if we take the transformation $\left\vert \varphi _{\lambda }^{\rho }(%
\mathbf{k})\right\rangle \rightarrow e^{i\chi \left( \mathbf{k}\right)
}\left\vert \varphi _{\lambda }^{\rho }(\mathbf{k})\right\rangle $ and $%
\left\vert \eta _{\lambda }^{\rho }(\mathbf{k})\right\rangle \rightarrow
e^{i\chi \left( \mathbf{k}\right) }\left\vert \eta _{\lambda }^{\rho }(%
\mathbf{k})\right\rangle $, the biorthonormal relation still holds; in
contrast, we have an additional term $i\nabla _{\mathbf{k}}\chi \left(
\mathbf{k}\right) $ in $\left( \overrightarrow{\mathcal{A}}_{\lambda }^{\rho
}\right) _{\mathrm{LR}}$.

The Berry curvature has the form
\begin{eqnarray}
\left( \overrightarrow{\mathcal{B}}_{\lambda }^{\rho }\right) _{\mathrm{LR}}
&=&\nabla _{\mathbf{k}}\times \overrightarrow{\mathcal{A}}_{\lambda }^{\rho }
\notag  \label{B} \\
&=&\mathbf{B}_{\lambda }^{\rho }+\nabla _{\mathbf{k}}\times \left[ \mathbf{N}%
_{\lambda }^{\rho }(\mathbf{k})\right] _{\mathrm{LR}},
\end{eqnarray}%
where $\mathbf{B}_{\lambda }^{\rho }=\nabla _{\mathbf{k}}\times \mathbf{A}%
_{\lambda }^{\rho }$. Equation (\ref{B}) means that the Berry curvature of $%
\mathcal{H}(\mathbf{k})$ is complex or real for real or imaginary $\epsilon
_{\lambda }^{\rho }(\mathbf{k})$. The additional term $\nabla _{\mathbf{k}%
}\times \left[ \mathbf{N}_{\lambda }^{\rho }(\mathbf{k})\right] _{\mathrm{LR}%
}$ yields zero integration over the Brillouin zone as we have shown in the
\textrm{RR} case,\ which means that the Chern numbers of $H(\mathbf{k})$ and
$\mathcal{H}(\mathbf{k})$ are identical
\begin{eqnarray}
\left( c_{\lambda }^{\rho }\right) _{\mathrm{LR}} &=&\frac{1}{2\pi }%
\oint\nolimits_{\mathrm{BZ}}\left( \overrightarrow{\mathcal{B}}_{\lambda
}^{\rho }\right) _{\mathrm{LR}}d^{2}\mathbf{k}  \notag  \label{c} \\
&=&\frac{1}{2\pi }\oint\nolimits_{\mathrm{BZ}}\mathbf{B}_{\lambda }^{\rho
}d^{2}\mathbf{k}.
\end{eqnarray}%
Integral in Eq. (\ref{c}) is under the assumption that EPs are absent in the
Brillouin zone, since the biorthonormal basis does not exist at EPs.

Besides the \textrm{RR} and \textrm{LR} definitions, one can also define the
\textrm{RL} and \textrm{LL} Berry connections and Berry curvatures. For the
Hermitian and corresponding non-Hermitian topological systems we concerned,
the \textrm{RL} and \textrm{LL }Berry connections are
\begin{widetext}
\begin{eqnarray}
\left( \overrightarrow{\mathcal{A}}_{\lambda }^{\rho }\right) _{\mathrm{RL}}
&=&i\left\langle \varphi _{\lambda }^{\rho }(\mathbf{k})\right\vert \nabla
_{ \mathbf{k}}\left\vert \eta _{\lambda }^{\rho }(\mathbf{k})\right\rangle =%
\mathbf{A}_{\lambda }^{\rho }-\frac{i\gamma }{\epsilon _{\lambda }^{\rho }(
\mathbf{k})}\left\langle \phi _{\lambda }^{-\rho }(\mathbf{k})\right\vert
i\nabla _{\mathbf{k}}\left\vert \phi _{\lambda }^{\rho }(\mathbf{k}%
)\right\rangle , \\
\left( \overrightarrow{\mathcal{A}}_{\lambda }^{\rho }\right) _{\mathrm{LL}}
&=&i\left\langle \eta _{\lambda }^{\rho }(\mathbf{k})\right\vert \nabla _{
\mathbf{k}}\left\vert \eta _{\lambda }^{\rho }(\mathbf{k})\right\rangle
=\left\{
\begin{array}{ll}
\mathbf{A}_{\lambda }^{\rho }+\frac{1}{2}\nabla _{\mathbf{k}}\vartheta , &
\text{real }\epsilon _{\lambda }^{\rho }(\mathbf{k})\\
\mathbf{A}_{\lambda }^{\rho }-\epsilon _{\lambda }^{\rho }(\mathbf{k}%
)\left\langle \phi _{\lambda }^{-\rho }(\mathbf{k})\right\vert i\nabla _{
\mathbf{k}}\left\vert \phi _{\lambda }^{\rho }(\mathbf{k})\right\rangle
/\left( i\gamma \right) , & \text{imaginary }\epsilon _{\lambda }^{\rho }(%
\mathbf{k}) \\
\mathbf{A}_{\lambda }^{\rho }, & \text{at }\mathrm{EPs}%
\end{array}%
\right. ,
\end{eqnarray}%
\end{widetext}
and the four definitions of the Chern number are identical for separated
bands (i.e., in the absence of EPs) \cite{Shen}.

\subsection{Details of the 1D comb lattice model}

\setcounter{equation}{0} \renewcommand{\theequation}{D\arabic{equation}} %
\setcounter{figure}{0} \renewcommand{\thefigure}{D\arabic{figure}}

\subsubsection{Model and energy bands}

The non-Hermitian Hamiltonian of the one-dimensional comb lattice model
reads
\begin{equation}
\mathcal{H}=H+i\gamma \sum_{j=1}^{2N}(a_{j}^{\dagger }a_{j}-b_{j}^{\dagger
}b_{j}),
\end{equation}%
which is generated from the Hermitian Hamiltonian
\begin{eqnarray}
H &=&\sum_{j=1}^{N}[J\left( 1-\delta \right) a_{2j}^{\dagger
}b_{2j-1}+J\left( 1+\delta \right) a_{2j}^{\dagger }b_{2j+1}]  \notag \\
&&+\sum_{j=1}^{N}\kappa _{-}a_{2j-1}^{\dagger }b_{2j-1}+\sum_{j=1}^{N}\kappa
_{+}a_{2j}^{\dagger }b_{2j}+\mathrm{H.c.},  \label{H}
\end{eqnarray}%
under the periodic boundary condition $b_{2N+1}=b_{1}$, and the system
parameters are $\delta =\delta _{0}+R\cos \theta $ and $\kappa _{\pm
}=\kappa _{0}\pm (1/2)R\sin \theta $ (set $\kappa \equiv \kappa _{+}-\kappa
_{-}=R\sin \theta $). We refer to the Hamiltonian with periodic boundary
condition as the bulk Hamiltonian, and the edge Hamiltonian is the
Hamiltonian under open boundary condition. Taking the Fourier transformation
\begin{equation}
\left(
\begin{array}{c}
a_{2j} \\
a_{2j-1} \\
b_{2j-1} \\
b_{2j}%
\end{array}%
\right) =\frac{1}{\sqrt{N}}\sum_{k}e^{ikj}\left(
\begin{array}{c}
a_{k,+} \\
a_{k,-} \\
b_{k,+} \\
b_{k,-}%
\end{array}%
\right) ,
\end{equation}%
we obtain
\begin{eqnarray}
H &=&\sum_{k}\alpha _{k}^{\dag }H(k)\alpha _{k}, \\
\mathcal{H} &=&\sum_{k}\alpha _{k}^{\dag }\mathcal{H}(k)\alpha _{k},
\end{eqnarray}%
where $\alpha _{k}^{\dag }=\left( a_{k,+}^{\dagger },a_{k,-}^{\dagger
},b_{k,+}^{\dagger },b_{k,-}^{\dagger }\right) $, and the $4\times 4$ matrix
$\mathcal{H}(k)$ and $H(k)$ has the form
\begin{eqnarray}
\mathcal{H}(k) &=&H(k)+i\gamma \sigma _{z}\otimes I_{2}  \notag \\
&=&\left(
\begin{array}{cccc}
i\gamma & 0 & \mu _{k} & \kappa _{+} \\
0 & i\gamma & \kappa _{-} & 0 \\
\mu _{-k} & \kappa _{-} & -i\gamma & 0 \\
\kappa _{+} & 0 & 0 & -i\gamma%
\end{array}%
\right) ,
\end{eqnarray}%
with $\mu _{k}=J\left( 1-\delta \right) +J\left( 1+\delta \right) e^{ik}$, $%
k_{m}=2\pi m/N$, $(m=1,2,...,N)$. The eigenstates of $H(k)$ has the form%
\begin{equation}
|\phi _{\lambda }^{\rho }(k)\rangle =\frac{1}{\Omega _{\lambda }^{\rho }(k)}%
\left(
\begin{array}{c}
\varepsilon _{\lambda }^{\rho }(k)\kappa _{-}\mu _{k} \\
\varepsilon _{\lambda }^{-\rho }(k)[\varepsilon _{-\lambda }^{\rho
}(k)^{2}-\kappa _{-}^{2}] \\
\kappa _{-}[\varepsilon _{\lambda }^{\rho }(k)^{2}-\kappa _{+}^{2}] \\
\kappa _{+}\kappa _{-}\mu _{k}%
\end{array}%
\right) ,
\end{equation}%
where $\Omega _{\lambda }^{\rho }(k)=\sqrt{2}\kappa _{-}\varepsilon
_{\lambda }^{\rho }(k)^{-1}\{[\varepsilon _{\lambda }^{\rho }(k)^{4}-\left(
\kappa _{+}\kappa _{-}\right) ^{2}][\varepsilon _{\lambda }^{\rho
}(k)^{2}-\left( \kappa _{+}\right) ^{2}]\}^{1/2}$ is the normalization
factor and $\left( \rho ,\lambda =\pm \right) $. The corresponding
eigenvalue is
\begin{equation}
\varepsilon _{\lambda }^{\rho }(k)=\rho \{\Upsilon _{k}+\lambda \lbrack
\Upsilon _{k}^{2}-\left( \kappa _{+}\kappa _{-}\right) ^{2}]^{\frac{1}{2}%
}\}^{\frac{1}{2}},
\end{equation}%
with $\Upsilon _{k}=(\left\vert \mu _{k}\right\vert ^{2}+\kappa
_{+}^{2}+\kappa _{-}^{2})/2$.

\begin{figure*}[bht]
\includegraphics[ bb=0 0 515 330, width=18.0 cm, clip]{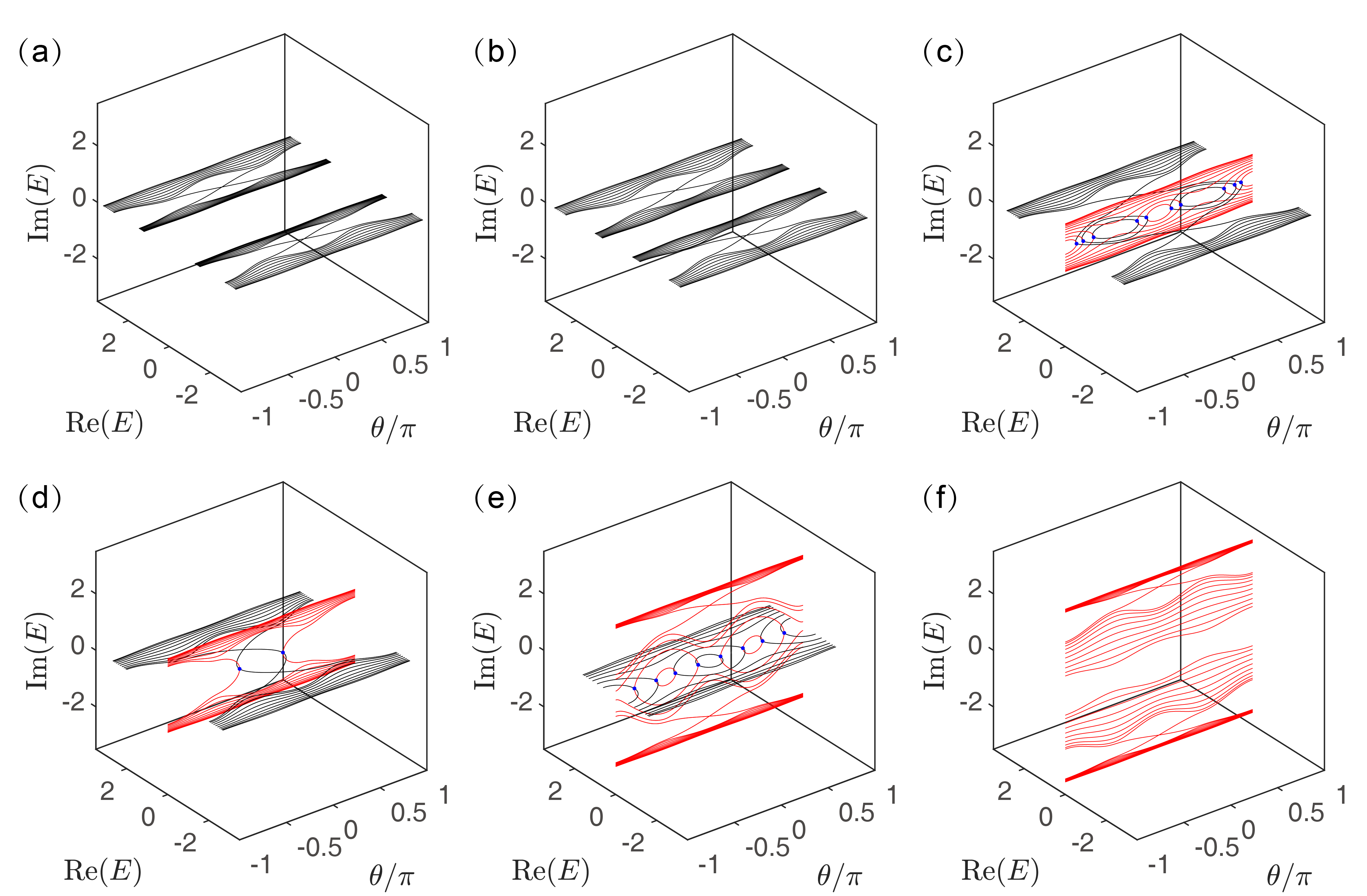}
\caption{Energy bands of the edge Hamiltonian for $\protect\delta_0=0$ at
(a) $\protect\gamma=0$, (b) $\protect\gamma=1.0$, (c) $\protect\gamma=1.5$,
(d) $\protect\gamma=1.8$, (e) $\protect\gamma=2.8$, (f) $\protect\gamma=3.3$
. The real (imaginary) part is in black (red), the blue dots are EPs. Other
parameters are $J=1$, $\protect\kappa_0=2$, $R=0.6$, and $N=10$.}
\label{figS2}
\end{figure*}

The energy bands are depicted in Fig.~\ref{figS2} at various $\gamma$ as the
supplementary of Figs.~2(a) and~2(b) in the main text.

\subsubsection{Zak phase}

In the condition of $\kappa _{+}=\kappa _{-}=\kappa _{0}$ ($\theta =0$), the
eigenstates of $\mathcal{H}(k)$ are
\begin{equation}
\left\vert \varphi _{\lambda }^{\rho }(k)\right\rangle =M_{\lambda }^{\rho
}\left( k,\gamma \right) \left\vert \phi _{\lambda }^{\rho }(k)\right\rangle
,
\end{equation}%
where
\begin{equation}
|\phi _{\lambda }^{\rho }(k)\rangle =\frac{1}{\Omega _{\lambda }^{\rho }(k)}%
\left(
\begin{array}{c}
\varepsilon _{\lambda }^{\rho }(k)\kappa _{0}\mu _{k} \\
\varepsilon _{\lambda }^{-\rho }(k)[\varepsilon _{-\lambda }^{\rho
}(k)^{2}-\kappa _{0}^{2}] \\
\kappa _{0}[\varepsilon _{\lambda }^{\rho }(k)^{2}-\kappa _{0}^{2}] \\
\kappa _{0}^{2}\mu _{k}%
\end{array}%
\right) ,
\end{equation}%
are the eigenstates of $H(k)$, and
\begin{eqnarray}
M_{\lambda }^{\rho }\left( k,\gamma \right) &=&\sqrt{\frac{2}{%
1+e^{2i\vartheta }}}\left(
\begin{array}{cccc}
e^{i\vartheta } & 0 & 0 & 0 \\
0 & e^{i\vartheta } & 0 & 0 \\
0 & 0 & 1 & 0 \\
0 & 0 & 0 & 1%
\end{array}%
\right) , \\
\vartheta &=&\arctan \left[ \gamma /\epsilon _{\lambda }^{\rho }\left(
k\right) \right] .
\end{eqnarray}

Similarly, the eigenstates of $\mathcal{H}^{\dagger }(k)$ are
\begin{equation}
\left\vert \eta _{\lambda }^{\rho }(k)\right\rangle =\left[ M_{\lambda
}^{\rho }\left( k,\gamma \right) \right] ^{\dagger }\left\vert \phi
_{\lambda }^{\rho }\right\rangle .
\end{equation}

By definition, the Berry connection of the non-Hermitian system reads
\begin{equation}
\mathcal{A}_{\lambda }^{\rho }=i\left\langle \eta _{\lambda }^{\rho
}(k)\right\vert \partial _{k}\left\vert \varphi _{\lambda }^{\rho
}(k)\right\rangle ,
\end{equation}%
in which, the right and left eigenstates can be written as
\begin{eqnarray}
\left\vert \varphi _{\lambda }^{\rho }(k)\right\rangle &=&S\left( \delta
\right) M_{\lambda }^{\rho }\left( k,\gamma \right) \left\vert \Phi
_{\lambda }^{\rho }(k)\right\rangle , \\
\left\vert \eta _{\lambda }^{\rho }(k)\right\rangle &=&S\left( \delta
\right) \left[ M_{\lambda }^{\rho }\left( k,\gamma \right) \right] ^{\dagger
}\left\vert \Phi _{\lambda }^{\rho }\right\rangle ,
\end{eqnarray}%
with
\begin{equation}
S\left( \delta \right) =\left(
\begin{array}{cccc}
\mu _{k} & 0 & 0 & 0 \\
0 & 1 & 0 & 0 \\
0 & 0 & 1 & 0 \\
0 & 0 & 0 & \mu _{k}%
\end{array}%
\right) ,
\end{equation}%
and
\begin{equation}
\left\vert \Phi _{\lambda }^{\rho }(k)\right\rangle =\frac{1}{\Omega
_{\lambda }^{\rho }(k)}\left(
\begin{array}{c}
\varepsilon _{\lambda }^{\rho }(k)\kappa _{0} \\
\varepsilon _{\lambda }^{-\rho }(k)\left[ \varepsilon _{-\lambda }^{\rho
}(k)^{2}-\kappa _{0}^{2}\right] \\
\kappa _{0}\left[ \varepsilon _{\lambda }^{\rho }(k)^{2}-\kappa _{0}^{2}%
\right] \\
\kappa _{0}^{2}%
\end{array}%
\right) .
\end{equation}%
Then the Zak phase $\mathcal{Z}_{\lambda }^{\rho }\left( \delta \right)
=\int_{-\pi }^{\pi }\mathcal{A}_{\lambda }^{\rho }dk$,%
\begin{widetext}
\begin{eqnarray}
\mathcal{Z}_{\lambda }^{\rho }\left( \delta \right) &=&i\int_{-\pi }^{\pi
}\left\langle \Phi _{\lambda }^{\rho }(k)\right\vert M_{\lambda }^{\rho
}\left( k,\gamma \right) S^{\dagger }\left( \delta \right) \partial _{k}%
\left[ S\left( \delta \right) M_{\lambda }^{\rho }\left( k,\gamma \right)
\left\vert \Phi _{\lambda }^{\rho }(k)\right\rangle \right] dk  \notag \\
&=&i\int_{-\pi }^{\pi }\left\langle \Phi _{\lambda }^{\rho }(k)\right\vert
M_{\lambda }^{\rho }\left( k,\gamma \right) S^{\dagger }\left( \delta
\right) \left[ \partial _{k}S\left( \delta \right) \right] M_{\lambda
}^{\rho }\left( k,\gamma \right) \left\vert \Phi _{\lambda }^{\rho
}(k)\right\rangle dk  \notag \\
&&+i\int_{-\pi }^{\pi }\left\langle \Phi _{\lambda }^{\rho }(k)\right\vert
M_{\lambda }^{\rho }\left( k,\gamma \right) S^{\dagger }\left( \delta
\right) S\left( \delta \right) \partial _{k}\left[ M_{\lambda }^{\rho
}\left( k,\gamma \right) \left\vert \Phi _{\lambda }^{\rho }(k)\right\rangle %
\right] dk.
\end{eqnarray}%
We note that $\left\vert \mu _{k}\left( \delta \right) \right\vert
^{2}=\left\vert \mu _{k}\left( -\delta \right) \right\vert ^{2}=4J^{2}\left[
\cos ^{2}\left( k/2\right) +\delta ^{2}\sin ^{2}\left( k/2\right) \right] $
and $\varepsilon _{\lambda }^{\rho }\left( k,\delta \right) =\varepsilon
_{\lambda }^{\rho }\left( k,-\delta \right) $, then
\begin{eqnarray}
S^{\dagger }\left( \delta \right) S\left( \delta \right)  &=&\left(
\begin{array}{cccc}
\left\vert \mu _{k}\left( \delta \right) \right\vert ^{2} & 0 & 0 & 0 \\
0 & 1 & 0 & 0 \\
0 & 0 & 1 & 0 \\
0 & 0 & 0 & \left\vert \mu _{k}\left( \delta \right) \right\vert ^{2}%
\end{array}%
\right) =S^{\dagger }\left( \left\vert \delta \right\vert \right) S\left(
\left\vert \delta \right\vert \right) , \\
\varepsilon _{\lambda }^{\rho }\left( k,\delta \right)  &=&\varepsilon
_{\lambda }^{\rho }\left( k,\left\vert \delta \right\vert \right) ,
\end{eqnarray}%
which means $\left\langle \Phi _{\lambda }^{\rho }\left( k\right)
\right\vert M_{\lambda }^{\rho }\left( k,\gamma \right) S^{\dagger }\left(
\delta \right) S\left( \delta \right) \partial _{k}\left[ M_{\lambda }^{\rho
}\left( k,\gamma \right) \left\vert \Phi _{\lambda }^{\rho }\left( k\right)
\right\rangle \right] $ is the function of $\left\vert \delta \right\vert $,
so we have
\begin{eqnarray}
\mathcal{Z}_{\lambda }^{\rho }\left( \delta \right) -\mathcal{Z}_{\lambda
}^{\rho }\left( -\delta \right)
=i\int_{-\pi }^{\pi }\left\langle \Phi _{\lambda }^{\rho }\right\vert
M_{\lambda }^{\rho }\left( \gamma \right) \left[ S^{\dagger }\left( \delta
\right) \partial _{k}S\left( \delta \right) -S^{\dagger }\left( -\delta
\right) \partial _{k}S\left( -\delta \right) \right] M_{\lambda }^{\rho
}\left( \gamma \right) \left\vert \Phi _{\lambda }^{\rho }\right\rangle dk.
\end{eqnarray}%
Direct derivation shows that
\begin{eqnarray}
\mathcal{Z}_{\lambda }^{\rho }\left( \delta \right) -\mathcal{Z}_{\lambda
}^{\rho }\left( -\delta \right)
=-8\delta J^{2}\kappa _{0}^{2}\int_{-\pi }^{\pi }\frac{\varepsilon
_{\lambda }^{\rho }(k)^{2}e^{2i\vartheta }+\kappa _{0}^{2}}{\Omega _{\lambda
}^{\rho }(k)^{2}\left( 1+e^{2i\vartheta }\right) }dk.
\end{eqnarray}%
Furthermore, using $e^{i\vartheta }=\left[ \epsilon _{\lambda }^{\rho
}(k)+i\gamma \right] /\varepsilon _{\lambda }^{\rho }(k),\epsilon _{\lambda
}^{\rho }(k)=\rho \sqrt{\varepsilon _{\lambda }^{\rho }(k)^{2}-\gamma ^{2}}%
,1+e^{2i\vartheta }=2e^{i\vartheta }\cos \vartheta ,$and $\tan \vartheta
=\gamma /\epsilon _{\lambda }^{\rho }(k)$, $\mathcal{Z}_{\lambda }^{\rho
}\left( \delta \right) -\mathcal{Z}_{\lambda }^{\rho }\left( -\delta \right)
$ is reduced to%
\begin{eqnarray}
\mathcal{Z}_{\lambda }^{\rho }\left( \delta \right) -\mathcal{Z}_{\lambda
}^{\rho }\left( -\delta \right)
=-\mathrm{sgn}\left( \delta \right) \pi -2i\gamma \rho \delta
J^{2}\int_{-\pi }^{\pi }\frac{[\varepsilon _{\lambda }^{\rho }(k)]^{2}}{%
\sqrt{\varepsilon _{\lambda }^{\rho }(k)^{2}-\gamma ^{2}}\left[ \varepsilon
_{\lambda }^{\rho }(k)^{4}-\kappa _{0}^{4}\right] }dk,
\end{eqnarray}%
\end{widetext}where the later term is imaginary and non-vanished for
non-Hermitian Hamiltonian; however, for $[\varepsilon _{\lambda }^{\rho
}(k)]^{2}=[\varepsilon _{\lambda }^{-\rho }(k)]^{2}$, the summation $\frac{1%
}{\pi }\sum_{\rho \lambda }\left[ \mathcal{Z}_{\lambda }^{\rho }\left(
\delta \right) -\mathcal{Z}_{\lambda }^{\rho }\left( -\delta \right) \right]
$ is an integer \cite{HJiang}. Zak phase is a physical interpretation of the
Chern number, since the adiabatic transport of particle is regarded as a
manifestation of Zak phase.

\subsubsection{Edge states}

The non-Hermitian Hamiltonian under open boundary condition is the edge
Hamiltonian
\begin{equation}
\mathcal{H}_{\mathrm{edge}}=\mathcal{H}-J\left( 1+\delta \right) \left(
a_{2N}^{\dagger }b_{1}+b_{1}^{\dagger }a_{2N}\right) .
\end{equation}%
The original Hermitian edge Hamiltonian possesses four edge states \cite%
{WRPRB}, from which we can obtain the corresponding edge states of the
non-Hermitian system by using the mapping method. The four edge states of $%
\mathcal{H}_{\mathrm{edge}}$ can be expressed as
\begin{equation}
\left\{
\begin{array}{c}
\left\vert \varphi _{\mathrm{R}}^{\pm }\right\rangle =\frac{1}{\sqrt{\Omega }%
}\sum\limits_{j=1}^{N}\varsigma ^{N-j}(e^{\pm i\vartheta _{\mathrm{R}%
}}a_{2j}^{\dagger }\pm b_{2j}^{\dagger })\left\vert \text{\textrm{vac}}%
\right\rangle \\
\left\vert \varphi _{\mathrm{L}}^{\pm }\right\rangle =\frac{1}{\sqrt{\Omega }%
}\sum\limits_{j=1}^{N}\varsigma ^{j-1}(e^{\pm i\vartheta _{\mathrm{L}%
}}a_{2j-1}^{\dagger }\pm b_{2j-1}^{\dagger })\left\vert \text{\textrm{vac}}%
\right\rangle%
\end{array}%
\right. ,
\end{equation}%
with eigenenergies
\begin{equation}
\left\{
\begin{array}{c}
\epsilon _{\mathrm{R}}^{\pm }=\pm \sqrt{\left( \kappa _{+}\right)
^{2}-\gamma ^{2}} \\
\epsilon _{\mathrm{L}}^{\pm }=\pm \sqrt{\left( \kappa _{-}\right)
^{2}-\gamma ^{2}}%
\end{array}%
\right. .
\end{equation}%
Here $\varsigma =(\delta -1)/(\delta +1)$, $\Omega =2(1-\varsigma
^{2N})/(1-\varsigma ^{2})$, $e^{\pm i\vartheta _{\mathrm{R}}}=(\epsilon _{%
\mathrm{R}}^{+}\pm i\gamma )/\kappa _{+}$ and $e^{\pm i\vartheta _{\mathrm{L}%
}}=(\epsilon _{\mathrm{L}}^{+}\pm i\gamma )/\kappa _{-}$.

\subsection{Topological charge pumping}

\setcounter{equation}{0} \renewcommand{\theequation}{E\arabic{equation}} %
\setcounter{figure}{0} \renewcommand{\thefigure}{E\arabic{figure}}

\begin{figure}[bht]
\includegraphics[ bb=0 0 575 230, width=8.7 cm, clip]{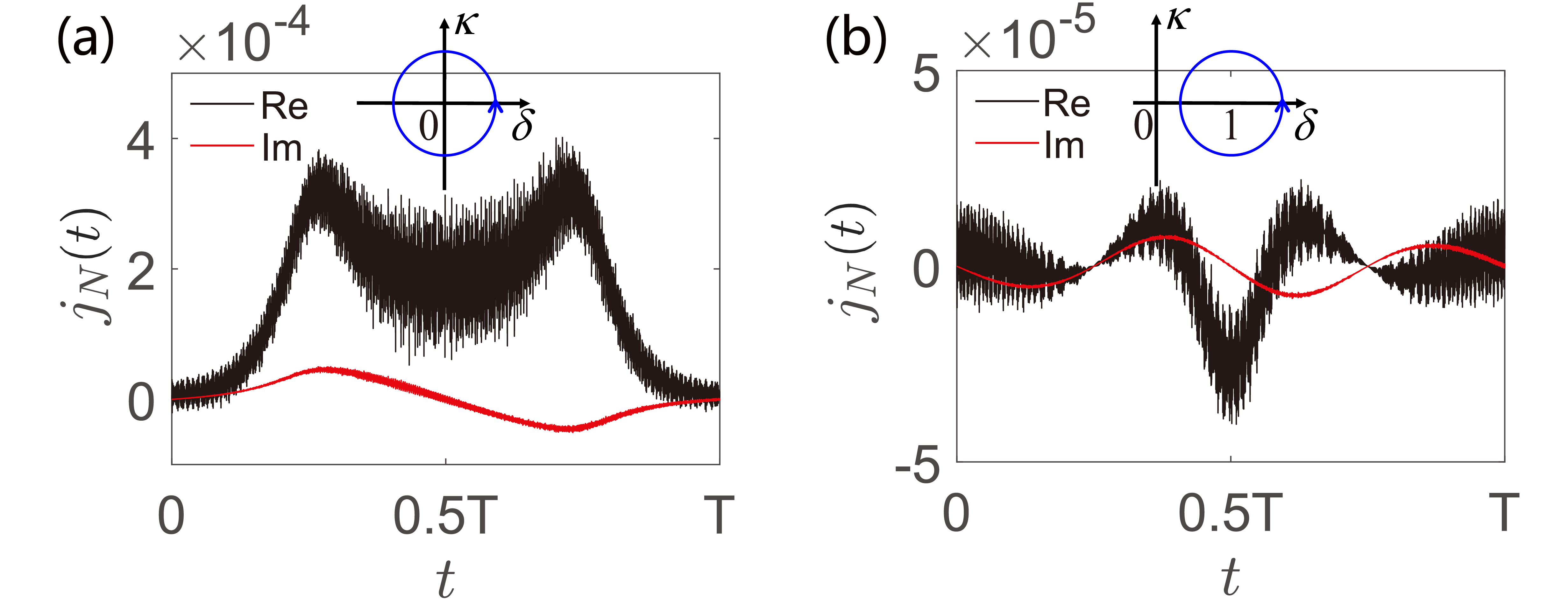}
\caption{Particle current $j_N(t)$ for the numerical simulations in Figs.
1(b) and 1(c) of the main text for $\mathcal{H}\left( t\right)$. Numerical
simulations are performed for (a) topological nontrivial phase $R>\left\vert
\protect\delta_{0}\right\vert$ at $\protect\delta_{0}=0$ and (b) topological
trivial phase $R<\left\vert\protect\delta_{0}\right\vert$ at $\protect\delta _{0}=1$. The speed of time evolution is $\protect\omega=0.001$ and the
period is $T=2\protect\pi\protect\omega^{-1}$. Other parameters are $\protect\gamma=0.5$, $J=1$, $R = 0.6$, and $N = 10$. Two quasi-adiabatic processes
are illustrated in the insets.} \label{figS3}
\end{figure}

\begin{figure*}[bth]
\includegraphics[ bb=0 0 580 115, width=17.8 cm, clip]{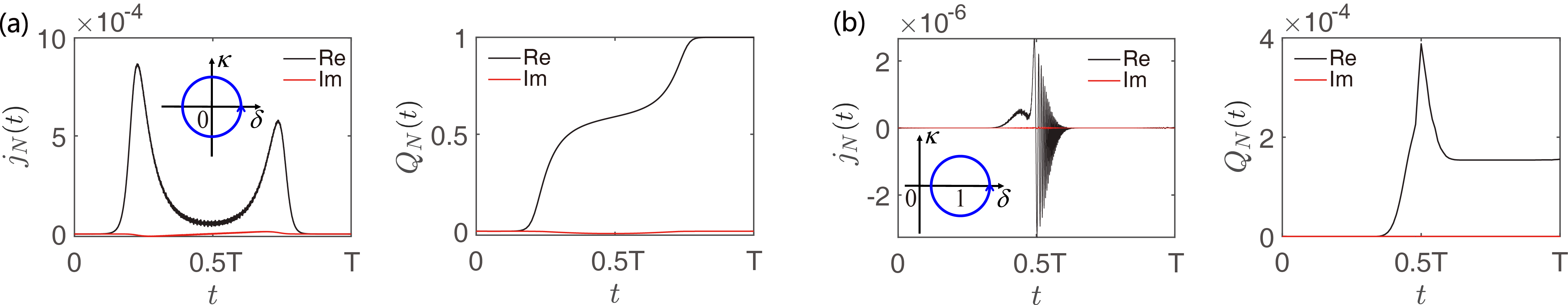}
\caption{Particle current $j_N(t)$ and topological charge pumping $Q_N(t)$
for the edge states of $\mathcal{H}_{\mathrm{edge}}(t)$. Numerical
simulations are performed for (a) topological nontrivial phase $R>\left\vert
\protect\delta_{0}\right\vert$ at $\protect\delta_{0}=0$ and (b) topological
trivial phase $R<\left\vert\protect\delta_{0}\right\vert$ at $\protect\delta %
_{0}=1$. The speed of time evolution is $\protect\omega=0.001$ and the
period is $T=2\protect\pi\protect\omega^{-1}$. Other parameters are $\protect%
\gamma=0.5$, $J=1$, $\protect\kappa_0=2$, $R = 0.6$, and $N = 10$. Two
quasi-adiabatic processes are illustrated in the insets.}
\label{figS4}
\end{figure*}

The non-Hermitian comb lattice Hamiltonian under periodic boundary condition
in the real space reads%
\begin{eqnarray}
\mathcal{H}\left( t\right) &=&\sum_{j=1}^{N}\{J\left[ 1-\delta \left(
t\right) \right] a_{2j}^{\dagger }b_{2j-1}+J\left[ 1+\delta \left( t\right) %
\right] a_{2j}^{\dagger }b_{2j+1}\}  \notag \\
&+&\sum_{j=1}^{N}\kappa _{-}\left( t\right) a_{2j-1}^{\dagger
}b_{2j-1}+\sum_{j=1}^{N}\kappa _{+}\left( t\right) a_{2j}^{\dagger }b_{2j}+%
\mathrm{H.c.}  \notag \\
&&+i\gamma \sum_{j=1}^{2N}(a_{j}^{\dagger }a_{j}-b_{j}^{\dagger }b_{j}).
\end{eqnarray}%
where $\delta \left( t\right) =\delta _{0}+R\cos \left( \omega t\right) $
and $\kappa _{\pm }\left( t\right) =\kappa _{0}\pm (1/2)R\sin \left( \omega
t\right) $. To examine how the scheme works in practice, we simulate the
quasi-adiabatic process by numerically computing the time evolution for a
finite system as discussed in the main text. The computation is performed by
using a uniform mesh in the time discretization for the time-dependent
Hamiltonian $\mathcal{H}\left( t\right) $. In order to demonstrate a
quasi-adiabatic process, we keep $f\left( t\right) =\left\vert \langle \bar{%
\eta}_{m}\left( t\right) \left\vert \varphi _{m}\left( t\right)
\right\rangle \right\vert \rightarrow 1$\ during the whole process by taking
sufficient small $\omega $, where $\left\vert \bar{\eta}_{m}\left( t\right)
\right\rangle $\ is the corresponding instantaneous eigenstate of $\mathcal{H%
}^{\dagger }\left( t\right) $. Figures \ref{figS3}(a) and \ref{figS3}(b)
depict the simulations of particle current for the topological nontrivial
and trivial phases, respectively. The corresponding total topological charge
pumping can be seen in Figs. 1(b) and 1(c) in the main text. We can see that
the imaginary parts of the currents yield zero integration in the interval $%
T $, and $Q_{N}\left( T\right) $\ are $1$ or $0$. The obtained dynamical
quantities are in close agreement with the Chern number.

The topological charge pumping can also be observed from the dynamics of
edge states in the quasi-adiabatic process of the lattice under open
boundary condition. The non-Hermitian edge Hamiltonian of the comb lattice
reads
\begin{equation}
\mathcal{H}_{\mathrm{edge}}(t)=\mathcal{H}(t)-J[1+\delta
(t)](a_{2N}^{\dagger }b_{1}+b_{1}^{\dagger }a_{2N}).
\end{equation}%
The biorthonormal current pumped by adiabatically varying $\theta \left(
t\right) $ across sites $a_{N}$\ and $b_{N-1}$ is defined as
\begin{equation}
j_{N}(t)=-i\langle \eta \left( t\right) |\{J\left[ 1-\delta (t)\right]
a_{N}^{\dagger }b_{N-1}-\mathrm{H.c.}\}|\varphi \left( t\right) \rangle .
\end{equation}%
To describe the process $|\varphi _{\mathrm{L}}^{\pm }\rangle \rightarrow
|\varphi _{\mathrm{R}}^{\pm }\rangle $, the accumulated Thouless charge
pumping passing the dimer $a_{N}$ and $b_{N-1}$\ during the interval $t$ is
\begin{equation}
Q_{N}(t)=\int_{0}^{t}j_{N}(t^{\prime })dt^{\prime }.
\end{equation}

Take $\theta \left( t\right) =\omega t$, $R>0$, $\omega \ll 1$, and the
initial state $|\varphi _{\mathrm{edge}}\left( 0\right) \rangle =|\varphi _{
\mathrm{L}}^{+}\rangle $. If $t$ varies from $0$ to $T=2\pi \omega ^{-1}$, $%
Q_{N}$\ should be $1$ \cite{WRPRB,WRPRA,KawabataPRB98}. We simulate the
quasi-adiabatic process by numerically computing the time evolution in a
finite system. In principle, for a given initial eigenstate $|\varphi _{
\mathrm{edge}}\left( 0\right) \rangle $, the evolved state under $\mathcal{H}%
_{\mathrm{edge}}\left( t\right) $ and $\mathcal{H}_{\mathrm{edge}}^{\dagger
}\left( t\right) $ is
\begin{equation}
\left\vert \varphi \left( t\right) \right\rangle =\mathcal{T}_{t}\{\exp
(-i\int_{0}^{t}\mathcal{H}_{\mathrm{edge}}\left( t^{\prime }\right) \mathrm{d%
}t^{\prime })\left\vert \varphi _{\mathrm{edge}}\left( 0\right)
\right\rangle \},
\end{equation}%
and
\begin{equation}
\left\vert \eta \left( t\right) \right\rangle =\mathcal{T}_{t}\{\exp
(-i\int_{0}^{t}\mathcal{H}_{\mathrm{edge}}^{\dagger }\left( t^{\prime
}\right) \mathrm{d}t^{\prime })\left\vert \eta _{\mathrm{edge}}\left(
0\right) \right\rangle \},
\end{equation}%
where $\mathcal{T}_{t}$ is the time ordering operator and $\left\vert \eta
_{ \mathrm{edge}}\left( 0\right) \right\rangle $ is the edge state for $%
\mathcal{H}_{\mathrm{edge}}^{\dagger }\left( 0\right) $ corresponding to $%
|\varphi _{\mathrm{edge}}\left( 0\right) \rangle $. In low speed limit $%
\omega \rightarrow 0$, we have $f\left( t\right) =\left\vert \langle \bar{
\eta}\left( t\right) \left\vert \varphi \left( t\right) \right\rangle
\right\vert \rightarrow 1$, where $\left\vert \bar{\eta}\left( t\right)
\right\rangle $\ is the corresponding instantaneous eigenstate of $\mathcal{H%
}_{\mathrm{edge}}^{\dagger }\left( t\right) $. The bulk-boundary
correspondence is that the topological charge pumping of an edge state for a
loop $L$ in the $\kappa $-$\delta $ plane equals to the Chern number.
Figures \ref{figS4}(a) and \ref{figS4}(b) depict the numerical simulations
of particle current and topological charge pumping for the edge states under
open boundary condition for the topological nontrivial and trivial phases in
the interval $T$, and the topological charge pumping is $1$ or $0$,
respectively.

\end{document}